\documentclass[showpacs,showkeys,preprintnumbers,eqsecnum,prd,twocolumn]{revtex4}  

\usepackage{graphicx}

\def\beq{\begin{equation}}
\def\eeq{\end{equation}}
\def\rmd{{\rm d}}

\begin{document}

\title{Deviation of quadrupolar bodies from geodesic motion in a Kerr spacetime}

\author{Donato Bini}
  \affiliation{
Istituto per le Applicazioni del Calcolo ``M. Picone,'' CNR, I-00185 Rome, Italy\\
ICRA, ``Sapienza" University of Rome, I-00185 Rome, Italy
}

\author{Andrea Geralico}
  \affiliation{Physics Department and ICRA, ``Sapienza" University of Rome, I-00185 Rome, Italy}

\date{\today}

\begin{abstract}
The deviation from geodesic motion of the world line of an extended body endowed with multipolar structure up to the mass quadrupole moment is studied in the Kerr background according to the Mathisson-Papapetrou-Dixon model.
The properties of the quadrupole tensor are clarified by identifying the relevant components which enter the equations of motion, leading to the definition of an effective quadrupole tensor sharing its own algebraic symmetries, but also obeying those implied by the Mathisson-Papapetrou-Dixon model itself.
The equations of motion are then solved analytically in the limit of small values of the characteristic length scales associated with the spin and quadrupole variables in comparison with the one associated with the background curvature and under special assumptions on body's structure and motion.
The resulting quasi-circular orbit is parametrized in a Keplerian-like form, so that temporal, radial and azimuthal eccentricities as well as semi-major axis, period and periastron advance are explicitly computed and expressed in terms of gauge-invariant variables in the weak field and slow motion limit.
A companion numerical study of the equations of motion is performed too.
\end{abstract}


\maketitle

\section{Introduction}

The motion of extended bodies in general relativity has been largely studied, starting from the pioneering works of Mathisson \cite{math37}, Papapetrou \cite{papa51,corpapa51}, Pirani \cite{pir56} and Tulczyjew \cite{tulc59}, which led to a self-consistent model describing the evolution of both linear and angular momentum of the body.
The inclusion of higher multipolar terms is due to Dixon \cite{dixon64,dixon69,dixon70,dixon73,dixon74}.
Ehlers and Rudolph \cite{ehlers77} then discussed in detail the quadrupolar structure of the body according to Dixon's formulation.

Recently, a renewed interest in the spinning body dynamics has been raised in the context of the two-body interaction, more and more accurately described within the post-Newtonian (PN) approximation of general relativity (see, e.g., Ref. \cite{Damour:2007nc} and references therein) as well as the associated \lq\lq effective-one-body" (EOB) approach (see Ref. \cite{Buonanno:1998gg} for the introduction of the EOB formalism and Ref. \cite{Damour:2008qf} for specific applications to the dynamics of two spinning black holes). 
Recently, Barausse, Racine and Buonanno \cite{Barausse:2009aa} derived a (constrained) Hamiltonian for a spinning particle in any curved background up to the linear order in spin, based on the work by Hanson and Regge \cite{Hanson:1974qy} and later developments by Porto and Rothstein \cite{porto}. 
Such an approach was then used to get an improved EOB Hamiltonian for spinning black hole binaries \cite{Barausse:2009xi}.
By construction, these results directly reduce to the dynamics of a spinning particle in a Kerr spacetime when specialized to the extreme mass-ratio limit, to linear order in the particle's spin. 
Therefore, studying the dynamics of a test body endowed with multipolar structure also serves as a consistency check of the Hamiltonian results and enables the extension to higher order in spin and to shape deformation described by the mass quadrupole moment \cite{hinderer}.
A number of related works by Buonanno and collaborators then followed over a period of few years \cite{Pan:2010hz,Blanchet:2011zv,Barausse:2011ys,Taracchini:2012ig,Buonanno:2012rv,Blanchet:2012sm,Blanchet:2012at, Pan:2013rra, Marsat:2013caa}.  Simultaneously, the Arnowitt-Deser-Misner (ADM) canonical Hamiltonian of two spinning objects was developed by Sch\"afer and collaborators to higher PN orders (see, e.g., Refs. \cite{Steinhoff:2008zr,Rothe:2010jj,Tessmer:2010hp,Wang:2011bt,Gopakumar:2011zz,Tessmer:2012xr,Tessmer:2013nk,Hartung:2013dza} and references therein).
Furthermore, corrections from the mass quadrupole are needed when the contributions quadratic in spin to the PN dynamics are taken into account (see, e.g., Refs. \cite{porto,steinhoff08,steinhoff09,hergt08}).

In this paper we study the dynamics of an extended body endowed with both spin and quadrupole moment in a Kerr spacetime according to the so called Mathisson-Papapetrou-Dixon (MPD) model. 
The motion is assumed to be confined on the equatorial plane, the spin vector of the body being aligned with the axis of rotation of the central object. 
Furthermore, the quadrupolar structure is simplified by assuming constant components of the quadrupole tensor with respect to the frame adapted to the body's generalized 4-momentum, due to the lack of associated evolution equations for higher multipoles in Dixon's model. The body is thus \lq\lq quasi-rigid,'' according to the definition of Ehlers and Rudolph \cite{ehlers77}.

The MPD equations of motion are then solved analytically in the limit of small values of the characteristic length scales associated with the spin and quadrupole variables with respect to the background curvature scale.
Initial conditions are chosen in such a way that the world line of the body and a timelike circular geodesic (taken as a reference world line) stem from a common spacetime point. 
The world line of the extended body thus deviates from the reference geodesic because of the combined effects of both the spin-curvature and quadrupole-curvature couplings. 
This approach allows an analytic discussion of the problem in complete generality in this limit, which is also compared with the numerical study of the full nonlinear equations.
The perturbative solution here found can be cast in a Keplerian-like form, by introducing the temporal, radial and azimuthal eccentricities of the orbit as well as the associated periods and frequencies.
We also compute the shift in the conserved energy and angular momentum due to both spin and quadrupole moment.

A generalized quasi-Keplerian representation of the orbit was introduced by Damour and Deruelle \cite{dam-derue1,dam-derue2} to parametrize the solution of 1PN accurate equations of motion for compact binaries in eccentric orbits. The generalization to the 2PN order in ADM coordinates is due to Damour, Sch\"afer, and Wex \cite{dam88,wex}.
The corresponding solution up to the third PN order has been obtained in Ref. \cite{gopa1} for two non-spinning compact objects, whereas the case of two spinning compact binaries taking into account the leading spin-orbit interaction in the dynamics is discussed in Ref. \cite{gopa2}. 
Such a Keplerian-like parametrization proves useful also in the present analysis.
In fact, the periastron advance and the orbital period are directly observable quantities.
Therefore, their measurements impose conditions on the experimental values of the gauge-invariant (i.e., coordinate-independent) conserved total energy and angular momentum, depending on a set of constant parameters, namely the spacetime mass and angular momentum as well as the spin and quadrupole moment of the extended body. 
In addition, taking the limit of weak field and slow motion of our perturbative solution allows us to distinguish among the different spin-spin interaction terms as well as the contribution due to the mass quadrupole moment of the body.

\section{Dixon's model and basic equations}

Consider an extended body endowed with structure up to the quadrupole, following the description due to Dixon \cite{dixon64,dixon69,dixon70,dixon73,dixon74}.
In the quadrupole approximation, Dixon's equations are
\begin{eqnarray}
\label{papcoreqs1}
\frac{{\rm D}P^{\mu}}{\rmd \tau} & = &
- \frac12 \, R^\mu{}_{\nu \alpha \beta} \, U^\nu \, S^{\alpha \beta}
-\frac16 \, \, J^{\alpha \beta \gamma \delta} \, \nabla^\mu R_{\alpha \beta \gamma \delta}
\nonumber\\
& \equiv & F^\mu_{\rm (spin)} + F^\mu_{\rm (quad)} \,,
\\
\label{papcoreqs2}
\frac{{\rm D}S^{\mu\nu}}{\rmd \tau} & = & 
2 \, P^{[\mu}U^{\nu]}+
\frac43 \, J^{\alpha \beta \gamma [\mu}R^{\nu]}{}_{\gamma \alpha \beta}
\nonumber\\
&\equiv & D^{\mu \nu}_{\rm (spin)} + D^{\mu \nu}_{\rm (quad)} \,,
\end{eqnarray}
where $P^{\mu}=m u^\mu$ (with $u \cdot u = -1$) is the total 4-momentum of the body with mass $m$, $S^{\mu \nu}$ is a (antisymmetric) spin tensor, $J^{\alpha\beta\gamma\delta}$ is the quadrupole tensor, and $U^\mu=\rmd z^\mu/\rmd\tau$ is the timelike unit tangent vector of the \lq\lq center of mass line'' (with parametric equations $x^\mu=z^\mu(\tau)$) used to make the multipole reduction, parametrized by the proper time $\tau$. 
Note that all tensors are defined {\it only} along the center of mass line, so that {\it only} the evolution along $U$ of such quantities is meaningful (analytic continuations off the reference world line would be arbitrary and unmotivated).
Furthermore, the spin force and torque in Eqs. (\ref{papcoreqs1}) and (\ref{papcoreqs2}) depend on both the tangent vector to the world line of multipole reduction $U$ and the generalized momentum unit vector $u$, but the same is not true for the quadrupolar force and torque, which only depend on the spacetime quantity $J^{\alpha\beta\gamma\delta}$ and the background geometry.

In order the model to be mathematically self-consistent certain additional conditions should be imposed. We adopt here the so called \lq\lq Tulczyjew supplementary conditions" \cite{tulc59,dixon64}
\beq
\label{tulczconds}
S^{\mu\nu}u{}_\nu=0\,.
\eeq
Consequently, the spin tensor can be fully represented by a spatial vector (with respect to $u$),
\beq
S(u)^\alpha=\frac12 \eta(u)^\alpha{}_{\beta\gamma}S^{\beta\gamma}=[{}^{*_{(u)}}S]^\alpha\,,
\eeq
where 
\beq
\eta(u)_{\alpha\beta\gamma}=\eta_{\mu\alpha\beta\gamma}u^\mu
\eeq
is the spatial (with respect to $u$) unit volume 3-form with $\eta_{\alpha\beta\gamma\delta}=\sqrt{-g} \epsilon_{\alpha\beta\gamma\delta}$ the unit volume 4-form and $\epsilon_{\alpha\beta\gamma\delta}$ ($\epsilon_{0123}=1$) the Levi-Civita alternating symbol. As standard, hereafter we denote the spacetime dual of a tensor (built up with $\eta_{\alpha\beta\gamma\delta}$) by a $^*$, whereas the 
spatial dual of a spatial tensor with respect to $u$ (built up with $\eta(u)_{\alpha\beta\gamma}$) by $^{*_{(u)}}$.
It is also useful to introduce the magnitude $s\ge0$ of the spin vector
\beq
\label{sinv}
s^2=S(u)^\beta S(u)_\beta = \frac12 S_{\mu\nu}S^{\mu\nu}\,, 
\eeq
which is in general not constant along the trajectory of the extended body. 

The tensor $J^{\alpha\beta\gamma\delta}$ has the same algebraic symmetries as the Riemann tensor, i.e.,
\beq
J^{\alpha\beta\gamma\delta}=J^{[\alpha\beta][\gamma\delta]}=J^{\gamma\delta\alpha\beta}\,,\qquad
J^{[\alpha\beta\gamma]\delta}=0\,,
\eeq
leading to 20 independent components.
Using standard spacetime splitting techniques associated with an arbitrary observer $\bar u$ it can be reduced to the form
\begin{eqnarray}
\label{deco_bar_u2}
J^{\alpha\beta\gamma\delta}&=&\eta(\bar u)^{\alpha \beta \mu}\eta(\bar u)^{\gamma\delta\nu}M(\bar u)_{\mu\nu}
+2\bar u^{[\alpha}W(\bar u)^{\beta]}{}_\sigma \eta(\bar u)^{\sigma \gamma\delta}\nonumber\\
&&+2\bar u^{[\gamma}W(\bar u)^{\delta]}{}_\sigma \eta(\bar u)^{\sigma \alpha\beta}
-4\bar u^{[\alpha}Q(\bar u)^{\beta][\gamma}\bar u^{\delta]}\,,\nonumber\\
\end{eqnarray}
where
\begin{eqnarray}
\label{Jsplit_baru}
Q(\bar u)_{\alpha\beta}&=&J_{\alpha\mu\beta\nu}\bar u^\mu \bar u^\nu\,,\nonumber\\ 
W(\bar u)_{\alpha\beta}&=&-[J^*]_{\alpha\mu\beta\nu}\bar u^\mu \bar u^\nu\,,\nonumber\\
M(\bar u)_{\alpha\beta}&=&[{}^*J^*]_{\alpha\mu\beta\nu}\bar u^\mu \bar u^\nu\,,
\end{eqnarray}
with $W(\bar u)_{\alpha\beta}$ trace-free.
When $\bar u=u$, i.e., when the observer's $4$-velocity is aligned with the body's $4$-momentum, the spatial tensors $Q(u)_{\alpha\beta}$ (6 independent components), $W(u)_{\alpha\beta}$ (8 independent components) and $M(u)_{\alpha\beta}$ (6 independent components), have an intrinsic meaning.

Notice that the representation (\ref{deco_bar_u2}) of $J$ is analogous to the standard $1+3$ representation of the Riemann tensor in terms of its electric ($E(\bar u)$), magnetic ($H(\bar u)$) and mixed ($F(\bar u)$) parts defined as
\begin{eqnarray}
E(\bar u)_{\alpha\beta}&=&R_{\alpha\mu\beta\nu}\bar u^\mu \bar u^\nu\,,\nonumber\\
H(\bar u)_{\alpha\beta}&=&-[R^*]_{\alpha\mu\beta\nu}\bar u^\mu \bar u^\nu\,,\nonumber\\
F(\bar u)_{\alpha\beta}&=&[{}^*R^*]_{\alpha\mu\beta\nu}\bar u^\mu \bar u^\nu\,.
\end{eqnarray}
In vacuum, where the mixed part $F(\bar u)_{\alpha\beta}=-E(\bar u)_{\alpha\beta}$, one has
\begin{eqnarray}
\label{rieman_deco}
R^{\alpha\beta\gamma\delta}\!\!&=&\!\!-\eta(\bar u)^{\alpha \beta \mu}\eta(\bar u)^{\gamma\delta\nu}E(\bar u)_{\mu\nu}
+2\bar u^{[\alpha}H(\bar u)^{\beta]}{}_\sigma \eta(\bar u)^{\sigma \gamma\delta}\nonumber\\
&&+2\bar u^{[\gamma}H(\bar u)^{\delta]}{}_\sigma \eta(\bar u)^{\sigma \alpha\beta}
-4\bar u^{[\alpha}E(\bar u)^{\beta][\gamma}\bar u^{\delta]}\,.\nonumber\\
\end{eqnarray}

\subsection{Symmetries of MPD equations and the \lq\lq effective'' quadrupole tensor}

The quadrupole tensor $J$ enters the MPD set of equations (\ref{papcoreqs1}) and (\ref{papcoreqs2}) only in contraction with the Riemann tensor and its covariant derivative, so that of its original 20 independent components only those obeying the symmetries of the equations will survive. 
We will show that this number is actually reduced to 10.

Let us introduce the spatial tensor $X(\bar u)$ defined by
\beq
M(\bar u)_{\alpha\beta}=Q(\bar u)_{\alpha\beta}+X(\bar u)_{\alpha\beta}\,,
\eeq
with
\beq
Q(\bar u)_{\alpha\beta}=[Q(\bar u)]^{\rm STF}{}_{\alpha\beta}+\frac13[{\rm Tr}\,Q(\bar u)]P(\bar u)_{\alpha\beta}\,,
\eeq
where $P(\bar u)^{\alpha}{}_{\beta}=\delta^{\alpha}{}_{\beta}+{\bar u}^{\alpha}{\bar u}_{\beta}$ projects orthogonally to ${\bar u}$ and STF stands for symmetric and trace-free.
The quadrupole tensor (\ref{deco_bar_u2}) can then be written as 
\begin{eqnarray}
\label{deco_bar_u3}
J^{\alpha\beta}{}_{\gamma\delta}&=& \delta^{\alpha\beta\mu}_{\gamma\delta\nu}[Q(\bar u)]^{\rm STF}{}_\mu{}^\nu+
\eta(\bar u)^{\alpha \beta \mu}\eta(\bar u)_{\gamma\delta}{}^{\nu}X(\bar u)_{\mu\nu}\nonumber\\
&& 
+2\bar u^{[\alpha}W(\bar u)^{\beta]}{}_\sigma \eta(\bar u)^{\sigma}{}_{ \gamma\delta}+2\bar u_{[\gamma}W(\bar u)_{\delta]}{}_\sigma \eta(\bar u)^{\sigma \alpha\beta}\nonumber\\
&&
+Z(\bar u)^{\alpha\beta}{}_{\gamma\delta}\,,
\end{eqnarray}
where 
\begin{eqnarray}
Z(\bar u)^{\alpha\beta}{}_{\gamma\delta}&=&
\frac13[{\rm Tr}\,Q(\bar u)]\left[
4{\bar u}^{[\alpha}P(\bar u)^{\beta]}{}_{[\gamma}{\bar u}_{\delta]}\right.\nonumber\\
&&\left.
+\eta({\bar u})^{\alpha\beta\mu}\eta({\bar u})_{\mu\gamma\delta}
\right]\,,
\end{eqnarray}
and the following relation has been used
\begin{eqnarray}
\label{STFrel}
\eta(\bar u)^{\alpha \beta \mu}\eta(\bar u)_{\gamma\delta}{}^{\nu}[Q(\bar u)]^{\rm STF}{}_{\mu\nu}&=&
4\bar u^{[\alpha}[Q(\bar u)]^{\rm STF}{}^{\beta]}{}_{[\gamma}\bar u_{\delta]}\nonumber\\
&&
+\delta^{\alpha\beta\mu}_{\gamma\delta\nu}[Q(\bar u)]^{\rm STF}{}_\mu{}^{\nu}\,.\nonumber\\
\end{eqnarray}
In the construction of both quadrupole force and torque defined in Eqs. (\ref{papcoreqs1}) and (\ref{papcoreqs2}) the terms involving $[Q(\bar u)]^{\rm STF}$ and $Z(\bar u)$ vanish for a Ricci-flat background.
Similarly, the contribution due to the pure-trace part of $X(\bar u)$ vanishes too, so that $X(\bar u)$ can be identified with its trace-free part.
Furthermore, after decomposing $W(\bar u)$ into its symmetric ($[W(\bar u)]^{\rm S}$) and antisymmetric ($[W(\bar u)]^{\rm A}$) parts one gets that the latter does not contribute. 
Therefore, without any loss of generality, we can assume the following decomposition for the quadrupole tensor
\begin{eqnarray}
\label{deco_bar_u4}
J^{\alpha\beta}{}_{\gamma\delta}&=&\eta(\bar u)^{\alpha \beta \mu}\eta(\bar u)_{\gamma\delta}{}^{\nu}[X(\bar u)]^{\rm STF}{}_{\mu\nu}\nonumber\\
&& 
+2\bar u^{[\alpha}[W(\bar u)]^{\rm STF}{}^{\beta]}{}_\sigma \eta(\bar u)^{\sigma}{}_{ \gamma\delta}\nonumber\\
&&
+2\bar u_{[\gamma}[W(\bar u)]^{\rm STF}{}_{\delta]}{}_\sigma \eta(\bar u)^{\sigma \alpha\beta}\,.
\end{eqnarray}
The number of effective components of $J$ is thus reduced to 10, being $X(\bar u)$ and $W(\bar u)$ both symmetric and trace-free spatial tensors, in agreement with the post-Newtonian treatment of extended bodies.

Notice that the property (\ref{STFrel}) holds also for $X(\bar u)$. In addition, the term $\delta^{\alpha\beta\mu}_{\gamma\delta\nu}[X(\bar u)]^{\rm STF}{}_\mu{}^\nu$ does not contribute when contracted with 
the Riemann tensor and its covariant derivative in a Ricci-flat background spacetime to give the quadrupole force and torque.
Therefore, one is allowed to replace the term $\eta(\bar u)^{\alpha \beta \mu}\eta(\bar u)_{\gamma\delta}{}^{\nu}[X(\bar u)]^{\rm STF}{}_{\mu\nu}$ in Eq. (\ref{deco_bar_u4}) by $4\bar u^{[\alpha}[Q(\bar u)]^{\rm STF}{}^{\beta]}{}_{[\gamma}\bar u_{\delta]}$, leading to the following definition of an \lq\lq effective'' quadrupole tensor (still denoted by $J$) which shares all the symmetries underlying the MPD equations
\begin{eqnarray}
\label{deco_bar_u5}
J^{\alpha\beta}{}_{\gamma\delta}&=&4\bar u^{[\alpha}[X(\bar u)]^{\rm STF}{}^{\beta]}{}_{[\gamma}\bar u_{\delta]}\nonumber\\
&& 
+2\bar u^{[\alpha}[W(\bar u)]^{\rm STF}{}^{\beta]}{}_\sigma \eta(\bar u)^{\sigma}{}_{ \gamma\delta}\nonumber\\
&&
+2\bar u_{[\gamma}[W(\bar u)]^{\rm STF}{}_{\delta]}{}_\sigma \eta(\bar u)^{\sigma \alpha\beta}\,.
\end{eqnarray}
The above discussion as well as the associated general expression for $J$ cannot be found in the reference works on this topic, thus representing by itself an original contribution to the general relativistic treatment of extended bodies.

\subsubsection*{Spin-induced quadrupole tensor}

One could also include in the definition of the quadrupole tensor also terms which are quadratic in spin.
For instance, one can consider the choice \cite{steinhoff,steinhoff2}
\beq
\label{Jspininduced}
J^{\alpha\beta\gamma\delta}=4{\bar u}^{[\alpha}X(\bar u)^{\beta][\gamma}{\bar u}^{\delta]}\,,
\eeq
with
\beq
X({\bar u})=\frac{C_Q}{m}[S^2]^{\rm STF}\,,
\eeq
where $C_Q$ is a constant and $[S^2]^{\rm STF}$ denotes the trace-free part of the square of the spin tensor, i.e., 
\begin{eqnarray}
[S^2]^{\rm STF}{}^{\alpha\beta}&=&S^{\alpha\mu}S_{\mu}{}^{\beta}-\frac13P(\bar u)^{\alpha\beta}S_{\rho\sigma}S^{\sigma\rho}\nonumber\\
&=&S(\bar u)^{\alpha}S(\bar u)^{\beta}-\frac13s^2P(\bar u)^{\alpha\beta}\nonumber\\
&=&[S(\bar u)\otimes S(\bar u)]^{\rm STF}{}^{\alpha\beta}\,,
\end{eqnarray}
where both the spin vector and the associated spin invariant have been used.
Special values of $C_Q$ have been given in Ref. \cite{steinhoff3}: for instance, in the case of a black hole one has $C_Q=1$ \cite{thorne}, whereas for neutron stars it depends on the equation of state and varies between 4.3 and 7.4 \cite{poisson}.  

Clearly, this choice is compatible with the vanishing of the magnetic part of the quadrupole tensor, i.e., no current of the mass quadrupole of the body is present in $J$ due to the spin.

\subsection{Papapetrou fields and conserved quantities in stationary and axisymmetric spacetimes}

When the background spacetime has Killing vectors, there are conserved quantities along the motion \cite{ehlers77}. For example, in the case of stationary axisymmetric spacetimes with coordinates adapted to the spacetime symmetries,  $\xi=\partial_t$ is the timelike Killing vector and $\eta=\partial_\phi$ is the azimuthal Killing vector. The corresponding conserved quantities are the total energy $E$ and the angular momentum $J$, namely
\begin{eqnarray}
\label{totalenergy}
E&=&-\xi_\alpha P^\alpha +\frac12 S^{\alpha\beta}F^{(t)}_{\alpha\beta}\,,\nonumber\\
J&=&\eta_\alpha P^\alpha -\frac12 S^{\alpha\beta}F^{(\phi)}_{\alpha\beta}\,,
\end{eqnarray}
where
\beq
F^{(t)}_{\alpha\beta}=\nabla_\beta \xi_\alpha=g_{t[\alpha,\beta]}\,, \quad
F^{(\phi)}_{\alpha\beta}=\nabla_\beta \eta_\alpha=g_{\phi[\alpha,\beta]}\,, 
\eeq
are the Papapetrou fields associated with the Killing vectors.
When considering circular orbits, with angular velocity $\zeta$ and tangent vector aligned with $k=\partial_t+\zeta\partial_\phi$, the above quantities can be combined to give 
\beq
E-\zeta J=-k_\alpha P^\alpha +\frac12 S^{\alpha\beta}k_{\alpha;\beta}\,,
\eeq
being 
\beq
F^{(t)}_{\alpha\beta}+\zeta F^{(\phi)}_{\alpha\beta}=k^\mu g_{\mu[\alpha,\beta]}=k_{[\alpha;\beta]}\,.
\eeq

\section{Dynamics of extended bodies in the equatorial plane of a Kerr spacetime}

The Kerr metric  in standard Boyer-Lindquist coordinates is given by
\begin{eqnarray}
\rmd s^2 &=& -\left(1-\frac{2Mr}{\Sigma}\right)\rmd t^2 -\frac{4aMr}{\Sigma}\sin^2\theta\rmd t\rmd\phi+ \frac{\Sigma}{\Delta}\rmd r^2\nonumber\\
&& 
+\Sigma\rmd \theta^2+\frac{(r^2+a^2)^2-\Delta a^2\sin^2\theta}{\Sigma}\sin^2 \theta \rmd \phi^2\ ,
\end{eqnarray}
where $\Delta=r^2-2Mr+a^2$ and $\Sigma=r^2+a^2\cos^2\theta$; here $a$ and $M$ are the specific angular momentum and total mass of the spacetime solution. The event horizon and inner horizon are located at $r_\pm=M\pm\sqrt{M^2-a^2}$.

Let us introduce the zero angular momentum observer (ZAMO) family of fiducial observers, with four velocity
\beq
\label{n}
n=N^{-1}(\partial_t-N^{\phi}\partial_\phi)\,;
\eeq
here $N=(-g^{tt})^{-1/2}$ and $N^{\phi}=g_{t\phi}/g_{\phi\phi}$ are the lapse and shift functions, respectively. A suitable orthonormal frame adapted to  ZAMOs is given by
\begin{eqnarray}
\label{ZAMO-frame}
e_{\hat t}&=&n , \,\quad
e_{\hat r}=\frac1{\sqrt{g_{rr}}}\partial_r, \,\nonumber\\
e_{\hat \theta}&=&\frac1{\sqrt{g_{\theta \theta }}}\partial_\theta, \,\quad
e_{\hat \phi}=\frac1{\sqrt{g_{\phi \phi }}}\partial_\phi ,
\end{eqnarray}
with dual 
\begin{eqnarray}
\label{ZAMO-frame-dual}
\omega^{{\hat t}}&=&N\rmd t\ , \quad 
\omega^{{\hat r}}=\sqrt{g_{rr}}\rmd r\ , \nonumber\\
\omega^{{\hat \theta}}&=& \sqrt{g_{\theta \theta }} \rmd \theta\ , \quad
\omega^{{\hat \phi}}=\sqrt{g_{\phi \phi }}(\rmd \phi+N^{\phi}\rmd t)\ .
\end{eqnarray}

The ZAMOs are accelerated with acceleration $a(n)=\nabla_n n$ and locally non-rotating, in the sense that their vorticity vector $\omega(n)^\alpha$ vanishes, but they have a nonzero expansion tensor $\theta(n)_{\alpha\beta}$; the latter, in turn, can be completely described by an expansion vector $\theta_{\hat \phi}(n)^\alpha=\theta(n)^\alpha{}_\beta\,{e_{\hat\phi}}^\beta$, that is
\beq
\label{exp_zamo}
\theta(n) = e_{\hat\phi}\otimes\theta_{\hat\phi}(n)
           +\theta_{\hat\phi}(n)\otimes e_{\hat\phi}
\,.
\eeq 
The trace of the expansion tensor $\theta(n)^\alpha{}_\alpha$ turns out to be zero.

The nonzero ZAMO kinematical quantities (i.e., acceleration and expansion) all belong to the $r$-$\theta$ 2-plane of the tangent space \cite{mfg,idcf1,idcf2,bjdf}, i.e.,
\begin{eqnarray}
\label{accexp}
a(n) & = & a(n)^{\hat r} e_{\hat r} + a(n)^{\hat\theta} e_{\hat\theta}\nonumber\\
 & = &\partial_{\hat r}(\ln N) e_{\hat r} + \partial_{\hat\theta}(\ln N)  e_{\hat\theta}
\,,
\nonumber\\
\theta_{\hat\phi}(n) 
& = & \theta_{\hat\phi}(n)^{\hat r}e_{\hat r} + \theta_{\hat\phi}(n)^{\hat\theta}e_{\hat \theta}\nonumber\\ 
  & = & -\frac{\sqrt{g_{\phi\phi}}}{2N}\,(\partial_{\hat r} N^\phi e_{\hat r} + \partial_{\hat\theta} N^\phi e_{\hat \theta})
\,.
\end{eqnarray}
In the static limit (as it is the case of a Schwarzschild field) $N^\phi\to0$ and the expansion vector $\theta_{\hat\phi}(n)$ vanishes. 

It is also useful to introduce the curvature vectors $\kappa(x^i,n)$ associated with the diagonal metric coefficients, i.e., explicitly,  
$\kappa(r,n)$, $\kappa(\theta,n)$ and $\kappa(\phi,n)$, defined by 
\begin{eqnarray}
\label{k_lie}
\kappa(x^i,n)
& = & \kappa(x^i,n)^{\hat r} e_{\hat r} + \kappa(x^i,n)^{\hat\theta} e_{\hat\theta}\nonumber\\
 & = & -[\partial_{\hat r}(\ln \sqrt{g_{ii}}) e_{\hat r} + \partial_{\hat\theta}(\ln \sqrt{g_{ii}})e_{\hat\theta}]
\,.
\end{eqnarray}
We will refer to $\kappa(\phi,n)^{\hat r}\equiv k_{\rm (lie)}$ as the Lie relative curvature (see Refs. \cite{idcf1,idcf2}, where such a notation was first introduced) when limiting to the case of equatorial orbits.

The ZAMO kinematical quantities as well as the non-vanishing frame components of the Riemann tensor are listed in Appendix A.

\subsection{Circular orbits on the equatorial plane}

The 4-velocity $U$ of a particle uniformly rotating on circular orbits
can be parametrized either by the (constant) angular velocity with respect to infinity $\zeta$ or equivalently by the (constant) linear velocity  $\nu$ with respect to the ZAMOs 
\beq
\label{orbita}
U=\Gamma [\partial_t +\zeta \partial_\phi ]=\gamma [e_{\hat t} +\nu e_{\hat \phi}], \qquad \gamma=(1-\nu^2)^{-1/2}\,,
\eeq
where
\begin{eqnarray}
\Gamma &=&\left[ N^2-g_{\phi\phi}(\zeta+N^{\phi})^2 \right]^{-1/2} = \frac{\gamma}{N}\,,\nonumber\\
\zeta&=&-N^{\phi}+\frac{N}{\sqrt{g_{\phi\phi}}}\,\nu \,.
\end{eqnarray}
Note that the azimuthal coordinate $\phi$ along the orbit depends on the coordinate time $t$ or proper time $\tau$ along that orbit according to 
\beq
\label{eq:phitau}
\phi-\phi_0  = \zeta (t-t_0) = \Omega (\tau-\tau_0) \,,\qquad
\Omega =\Gamma\zeta\,,
\eeq
defining the corresponding coordinate and proper time orbital angular velocities $\zeta$ and $\Omega$. 
It is useful to introduce a spacelike unit vector $\bar U$ within the Killing 2-plane which is orthogonal to $U$ given by
\beq
\label{barU}
\bar U=\bar \Gamma [\partial_t +\bar \zeta \partial_\phi ]=\gamma [\nu e_{\hat t} +e_{\hat \phi}]\ ,
\eeq
with
\beq
\bar \zeta=-\frac{g_{tt}+\zeta g_{t\phi}}{g_{t\phi}+\zeta g_{\phi\phi}}
  =-N^{\phi}+\frac{N}{\sqrt{g_{\phi\phi}}}\, \frac{1}{\nu}\ , \qquad 
\bar \Gamma=\Gamma\nu\ .
\eeq

We limit our analysis to  the equatorial plane ($\theta=\pi/2$) of the Kerr solution, where
\begin{eqnarray}
N&=&\left[\frac{r\Delta}{r^3+a^2r+2a^2M}\right]^{1/2}\,,\nonumber\\
N^\phi&=&-\frac{2aM}{r^3+a^2r+2a^2M}\,,
\end{eqnarray}
and $\Delta=N^2g_{\phi\phi}$.
As a convention, the physical (orthonormal) component along $-\partial_\theta$, perpendicular to the equatorial plane will be referred to as along the positive $z$-axis and will be indicated by $\hat z$, when necessary.

\subsubsection*{Circular geodesics}

The angular and linear velocities associated with co-rotating ($U_{{(\rm geo)}+}\equiv U_+$, to shorten notation) and counter-rotating ($U_{{(\rm geo)}-}\equiv U_-$) timelike circular geodesics are given by
\begin{eqnarray}
\zeta_{\pm}&=&\pm\frac{\zeta_K}{1\pm a\zeta_K}\,,\nonumber\\
\nu_\pm &=&\frac{r^2\zeta_\pm}{\sqrt{\Delta}}\left(1+\frac{a^2}{r^2}\mp2a\zeta_K\right)\,,
\end{eqnarray}
respectively, where $\zeta_K=\sqrt{M/r^3}$ is the Keplerian angular velocity in the static case.
Note that $\nu_\pm$ is not positively defined and in the Schwarzschild limit ($a=0$) it reduces to 
\beq
\lim_{a \to 0}\nu_\pm =\pm\sqrt{\frac{M}{r-2M}}\equiv \pm \nu_K \,.
\eeq  
The corresponding timelike conditions $|\nu_\pm|<1$ identify the allowed regions for the radial coordinate where co/counter-rotating geodesics exist: $r>r_{{(\rm geo)}\pm}$, where 
\beq
r_{{(\rm geo)}\pm}=2M\left\{1+\cos\left[\frac23\arccos\left(\pm\frac{a}{M}\right)\right]\right\}\ .
\eeq

Therefore, the unit tangent vector to the timelike circular geodesics $U_{(\rm geo)}$ has the following (contravariant as well as covariant, in which case we use  the qualifier $\flat$) form 
\begin{eqnarray}
\label{Ugeodef}
U_{(\rm geo)\, \pm}&\equiv&U_{\pm}=\Gamma_{\pm} [\partial_t +\zeta_{\pm} \partial_\phi ]=\gamma_{\pm} [e_{\hat t} +\nu_{\pm} e_{\hat \phi}]\,,
\nonumber\\
U_{(\rm geo)\, \pm}^\flat&=& -\tilde E_\pm\rmd t+\tilde L_\pm\rmd\phi\,,
\end{eqnarray}
where $\tilde E_\pm$ and $\tilde L_\pm$ are the energy and azimuthal angular momentum per unit mass of the particle, respectively, given by
\begin{eqnarray}
\tilde E_\pm&=&N\gamma_\pm\left(1+\frac{2aM}{r\sqrt{\Delta}}\nu_\pm\right)
=\frac{|\Omega_\pm|}{\zeta_K}\left(1-\frac{2M}{r}\pm a\zeta_K\right)\,,
\nonumber\\
\tilde L_\pm&=&\gamma_\pm\nu_\pm\sqrt{g_{\phi\phi}}
=\Omega_\pm r^2\left(1+\frac{a^2}{r^2}\mp2a\zeta_K\right)\,,
\end{eqnarray}
with $\Omega_\pm=\Gamma_\pm\zeta_\pm$. Therefore we find 
\beq
\tilde E_\pm-\zeta_\pm\tilde L_\pm=\frac{1}{\Gamma_\pm}\,,
\eeq
and the following relations hold
\begin{eqnarray}
\Gamma_\pm&=&\frac{\zeta_K}{|\zeta_\pm|}\left(1-\frac{3M}{r}\pm 2a\zeta_K\right)^{-1/2}\,,
\nonumber\\
\bar\Gamma_\pm&=&|\Omega_\pm|\frac{r^2}{\sqrt{\Delta}} \left(1+\frac{a^2}{r^2}\mp2a\zeta_K\right)\,,
\nonumber\\
\bar\zeta_\pm&=&\pm\frac{r\zeta_K}{M}\frac{1-{2M}/{r}\pm a\zeta_K}{1+{a^2}/{r^2}\mp2a\zeta_K}
=\frac{\tilde E_\pm}{\tilde L_\pm}\,.
\end{eqnarray}

Finally, the parametric equations of $U_{\pm}$ are then given by
\beq
\label{circgeos}
t_\pm=t_0+\Gamma_\pm \tau\,,\quad 
r=r_0\,,\quad  
\theta=\frac{\pi}{2}\,,\quad
\phi_\pm=\phi_0+\Omega_\pm \tau\,,
\eeq
where $t_0$, $r_0$ and $\phi_0$ are constants.

\subsection{Orbit of the extended body}

Let the world line of the extended body with unit tangent vector $U$ be confined on the equatorial plane and not circular in general, i.e.,  
\beq
\label{polarnu}
U=\gamma(U,n) [n+ \nu(U,n)]\,,
\eeq
with
\beq
\label{polarnu2}
\nu(U,n)\equiv \nu^{\hat r}e_{\hat r}+\nu^{\hat \phi}e_{\hat \phi} 
=  \nu (\cos \alpha e_{\hat r}+ \sin \alpha e_{\hat \phi})\,,
\eeq
where $\gamma(U,n)=1/\sqrt{1-||\nu(U,n)||^2}\equiv\gamma$ is the Lorentz factor and the abbreviated notation $\nu^{\hat a}\equiv\nu(U,n)^{\hat a}$ has been used.  Similarly $\nu \equiv ||\nu(U,n)||$ and $\alpha$ are the magnitude of the spatial velocity $\nu(U,n)$ and its
polar angle  measured clockwise from the positive $\phi$ direction in the $r$-$\phi$ tangent plane respectively, while $\hat\nu\equiv\hat \nu(U,n)$ is the associated unit vector.
Note that $\alpha=\pi/2$ corresponds to azimuthal motion with respect to the ZAMOs, while 
$\alpha=0,\pi$ correspond to (outward/inward) radial motion with respect to the ZAMOs.

A convenient adapted frame to $U$ is given by
\begin{eqnarray}
\label{U_frame}
E_1&\equiv& \hat \nu ^\perp=\sin\alpha  e_{\hat r}- \cos\alpha  e_{\hat \phi}\,,\nonumber\\
E_2&=&\gamma [\nu n+\hat \nu]\,,\quad
E_3=-e_{\hat \theta}\,.
\end{eqnarray}
  
A similar decomposition holds for the 4-momentum $P=mu$ for equatorial motion, i.e.,  
\beq
\label{polarnuu}
u=\gamma_u [n +\nu_u\hat \nu_u]\,, \qquad
\gamma_u=(1-\nu_u^2)^{-1/2}\,,
\eeq
with
\beq
\label{polarnuu2}
\hat \nu (u,n)\equiv \hat \nu_u=\cos\alpha_u e_{\hat r}+ \sin\alpha_u e_{\hat \phi}\,.
\eeq
An orthonormal frame adapted to $u\equiv e_0$ is then built with the spatial triad
\begin{eqnarray}
\label{uframe}
e_1&\equiv& \hat \nu_u^\perp=\sin\alpha_u e_{\hat r}- \cos\alpha_u e_{\hat \phi}\,,\nonumber\\
e_2&=&\gamma_u [\nu_u n +\hat \nu_u]\,,\quad
e_3=-e_{\hat \theta}\,.
\end{eqnarray}
The dual frame of $\{e_\alpha\}$ will be denoted by $\{\omega^\alpha\}$, with $\omega^0=-u^\flat$.

The projection of the spin tensor into the local rest space of $u$ defines the spin vector $S(u)$ (hereafter simply denoted by $S$, for short).
When decomposed with respect to the frame (\ref{ZAMO-frame}) adapted to $n$ the spin vector is then given by
\beq
S=S^{\hat t}e_{\hat t}+S^{\hat r}e_{\hat r}+S^{\hat \theta}e_{\hat \theta}+S^{\hat \phi}e_{\hat \phi}\,,
\eeq
with $S^{\hat t}=\nu_u[S^{\hat r}\cos\alpha_u+S^{\hat \phi}\sin\alpha_u]$ due to the supplementary conditions (\ref{tulczconds}).
When decomposed with respect to the frame (\ref{uframe}) adapted to $u$ it writes instead as
\beq
S=S^1e_1+S^2e_2+S^3e_3\,.
\eeq

\subsection{Setting the body's spin and quadrupole}

In the following we will consider the special case in which the spin vector is aligned along the $z$-axis, i.e.,
\beq
S=S^{\hat\theta}e_{\hat \theta}=se_{\hat z}=se_3\,.
\eeq
When decomposed with respect to the frame adapted to $u$, the spin force and torque defined in Eqs. (\ref{papcoreqs1}) and (\ref{papcoreqs2}) are thus given by
\beq
\label{fspinframeu}
F_{\rm (spin)}=F_{\rm (spin)}^0u+F_{\rm (spin)}^1e_1+F_{\rm (spin)}^2e_2\,,
\eeq
and 
\beq
\label{dspinframeu}
D_{\rm (spin)}=-\omega^0\wedge{\mathcal E}(u)_{\rm (spin)}\,,
\eeq
with
\beq
\label{dspinframeu2}
{\mathcal E}(u)_{\rm (spin)}={\mathcal E}(u)_{\rm (spin)}{}_1\omega^{1}+{\mathcal E}(u)_{\rm (spin)}{}_2\omega^{2}\,,
\eeq
respectively.
The explicit expressions for the components are listed in Appendix B. 

Furthermore, we will assume the quadrupole tensor having constant frame components with respect to the frame (\ref{uframe}) adapted to $u$ as the most natural and simplifying choice. 
According to the terminology introduced in Ref. \cite{ehlers77}, the extended body should be termed in this case as \lq\lq quasi-rigid.'' 
Other approaches (equally valid in the framework of the MPD model) assume the quadrupole tensor be directly related to the Riemann tensor, having the same symmetry properties.
However, none of these choices, even if very convenient from a computational point of view, has a transparent physical meaning ---at least a priori--- in the context of the MPD model, because the quadrupole tensor represents the matter content of the body, and cannot be specified at all by the background in which the body moves.

Let the quadrupole tensor be given by Eq. (\ref{deco_bar_u5}) with $\bar u=u$.
In order that the motion be confined on the equatorial plane we must require $X(u)_{13}=X(u)_{23}=0$ and $W(u)_{11}=W(u)_{22}=W(u)_{12}=0$.
The quadrupole force and torque with respect to the frame adapted to $u$ are then given by
\beq
\label{fquadframeu}
F_{\rm (quad)}=F_{\rm (quad)}^0u+F_{\rm (quad)}^1e_1+F_{\rm (quad)}^2e_2\,,
\eeq
and
\beq
\label{dquadframeu}
D_{\rm (quad)}=-\omega^0\wedge{\mathcal E}(u)_{\rm (quad)}+{}^{*_{(u)}}{\mathcal B}(u)_{\rm (quad)}\,,
\eeq
with
\begin{eqnarray}
\label{dquadframeu2}
{\mathcal E}(u)_{\rm (quad)}&=&{\mathcal E}(u)_{\rm (quad)}{}_1 \omega^1 +{\mathcal E}(u)_{\rm (quad)}{}_2 \omega^2\,,\nonumber\\
{\mathcal B}(u)_{\rm (quad)}&=&{\mathcal B}(u)_{\rm (quad)}{}_3 \omega^3\,,
\end{eqnarray}
so that ${\mathcal E}(u)_{\rm (quad)}\cdot {\mathcal B}(u)_{\rm (quad)}=0$, respectively.
To further simplify the description of the extended body we assume that the quadrupole tensor in the $u$-frame be represented by two independent components only, i.e., $X(u)_{11}$ and $X(u)_{22}$, with $X(u)_{33}=-X(u)_{11}-X(u)_{22}$, the remaining components being set equal to zero (namely, $W(u)_{13}=W(u)_{23}=X(u)_{12}=0$).
Such a condition can be relaxed, e.g., if one is interested in a fully numerical description of the problem.
The explicit expressions for the components of the quadrupole force and torque are listed in Appendix D. 

Note that the special choice $X(u)_{11}=X(u)_{22}=-C_Qs^2/(3m)$ covers the results of Ref. \cite{hinderer}.

\section{Deviation from a circular geodesic}
\label{pertsol}

In order to avoid backreaction effects, implicit in the MPD model is the requirement that the structure of the body should produce very small deviations from geodesic motion in the sense that the natural length scales associated with the body, i.e., the \lq\lq bare'' mass $m_0$, the spin length $|S^a|/m_0$ and the quadrupolar lengths $(|Q(u)_{ab}|/m_0)^{1/2}$, $(|W(u)_{ab}|/m_0)^{1/2}$ and $(|M(u)_{ab}|/m_0)^{1/2}$, must be small enough if compared with the length scale associated with the background curvature.
Therefore, it seems reasonable to introduce the conditions of ``small spin" and ``small quadrupole" from the very beginning, resulting in a simplified set of linearized differential equations which can be easily integrated.
It is useful to define the following dimensionless parameters 
\beq
{\hat s}=\frac{s}{m_0M}\,,\qquad
{\hat X}_{ab}=\frac{X(u)_{ab}}{m_0M^2}\,,
\eeq
associated with spin and quadrupole respectively, which will be taken to be much smaller than unity as smallness indicators (i.e., ${\hat s}\ll1$ and $|{\hat X}_{ab}|\ll1$).
Note that in this approximation scheme quantities which are linear in ${\hat s}$ will be considered as ``first order," whereas quantities which are linear in ${\hat X}_{ab}$ will be considered as ``second order." 
Therefore, the spin will contribute both to the first order and to the second, whereas the quadrupole to the second order only. 

Consider then a pair of world lines emanating from a common spacetime point,
one a geodesic with 4-velocity $U_{(\rm geo)}$, the other a world line of an extended body which deviates from the reference one because of the combined effects of geodesic deviation and both the spin-curvature and quadrupole-curvature couplings, with 4-velocity $U$. 
Solutions of the equation of motion can then be found in the general form
\begin{eqnarray}
\label{U_pert_v}
x^\alpha&=&x^\alpha_{\rm (geo)}+x^\alpha_{(1)}+x^\alpha_{(2)}\,, \nonumber\\
U&=&U_{\rm (geo)}+U_{(1)}+U_{(2)}\,,
\end{eqnarray}
where the subscripts indicate the order of approximation.
It is worth noting that that $U_{\rm (geo)}$ and $U$ are unit tangent vectors to different world lines, which are parametrized by different proper times: hence, one should use $\tau_{\rm (geo)}$ as the proper time parameter along $U_{\rm (geo)}$ and $\tau$ as the proper time parameter along $U$. 
However, recalling the definitions of the proper time parameter along these world lines
\beq
d\tau=-U_\alpha dx^\alpha\,,\qquad 
d\tau_{\rm (geo)}=-U_{\rm (geo)}{}_\alpha dx_{\rm (geo)}^\alpha
\eeq
and using the normalization condition $U\cdot U=-1$, one obtains that $\tau$ and $\tau_{\rm (geo)}$ can be identified to that order of approximation, i.e.,
\beq
\tau=\tau_{\rm (geo)}+O(3)\,.
\eeq
Therefore, although the two world lines are parametrized by different proper times, the latter are synchronized so that $\tau$ can be used unambiguously for that single proper time parametrization of both world lines.

Let us solve the system of MPD equations perturbatively, by assuming that $U$ be tangent to a geodesic circular orbit in the equatorial plane at radius $r=r_0$ for vanishing spin and quadrupole. 
The geodesic 4-velocity $U_{\rm (geo)}$ is thus given by Eq. (\ref{Ugeodef}), where all quantities are understood to be evaluated at $r=r_0$ and a positive (negative) sign corresponds to co-rotating (counter-rotating) orbits with respect to increasing values of the azimuthal coordinate $\phi$. 
The first and second order corrections to the 4-velocity $U$ of the extended body are then obtained by taking the expansion of the general form (\ref{polarnu})--(\ref{polarnu2}) according to 
\beq 
\nu=\nu_\pm+\nu_{(1)}+\nu_{(2)}\,,\quad
\alpha=\frac{\pi}{2}+\alpha_{(1)}+\alpha_{(2)}\,,
\eeq
or equivalently
\beq 
\nu^{\hat r}=\nu^{\hat r}_{(1)}+\nu^{\hat r}_{(2)}\,,\quad
\nu^{\hat \phi}=\nu_\pm+\nu^{\hat \phi}_{(1)}+\nu^{\hat \phi}_{(2)}\,,
\eeq
so that 
\begin{eqnarray}
\nu^{\hat r}_{(1)}&=&-\nu_\pm\alpha_{(1)}\,,\quad
\nu^{\hat r}_{(2)}=-\nu_{(1)}\alpha_{(1)}-\nu_\pm\alpha_{(2)}\,,\nonumber\\
\nu^{\hat \phi}_{(1)}&=&\nu_{(1)}\,,\quad
\nu^{\hat \phi}_{(2)}=-\frac12\nu_\pm\alpha_{(1)}^2+\nu_{(2)}\,.
\end{eqnarray}
We find
\beq
U_{(1)}=\gamma_\pm\left[\nu_{(1)}^{\hat r}e_{\hat r}+\left(\gamma_\pm\nu_{(1)}^{\hat \phi}-\frac{\nu_\pm}{\gamma_\pm}\frac{r_0}{\sqrt{\Delta}}k_{\rm (lie)}r_{(1)}\right){\bar U}_\pm\right]\,,
\eeq
and
\beq
U_{(2)}=\gamma_\pm\left[X U_{\rm (geo)}+Y {\bar U}_\pm+Z e_{\hat r}\right]\,,
\eeq
with
\begin{eqnarray}
X&=&\gamma_\pm\left[\sqrt{g_{rr}}k_{\rm (lie)}r_{(1)}\nu_{(1)}^{\hat r}+\frac{\gamma_\pm}2\left((\nu_{(1)}^{\hat r})^2+\gamma_\pm(\nu_{(1)}^{\hat \phi})^2\right)\right]\nonumber\\
&&
+\frac1{2\gamma_\pm}\left[g_{rr}(\nu_\pm^2k_{\rm (lie)}^2-4\zeta_K^2)+3\Omega_\pm^2\right]r_{(1)}^2
\,,\nonumber\\
Y&=&\gamma_\pm\left(\gamma_\pm^2\nu_\pm(\nu_{(1)}^{\hat \phi})^2+\nu_{(2)}^{\hat \phi}\right)
-\frac{\sqrt{g_{rr}}}{\gamma_\pm}(k_{\rm (lie)}\pm2\zeta_K)r_{(2)}\nonumber\\
&&
-\gamma_\pm\sqrt{g_{rr}}[\nu_\pm(k_{\rm (lie)}\pm2\zeta_K)-k_{\rm (lie)}]r_{(1)}\nu_{(1)}^{\hat \phi}\nonumber\\
&&
+\frac{g_{rr}}{2\gamma_\pm\nu_\pm}\left\{
2[\nu_\pm(k_{\rm (lie)}\pm2\zeta_K)-k_{\rm (lie)}]^2\right.\nonumber\\
&&
\left.-2k_{\rm (lie)}(k_{\rm (lie)}\mp\nu_\pm\zeta_K)+\nu_\pm\kappa(r,n)^{\hat r}(k_{\rm (lie)}\pm2\zeta_K)\right.\nonumber\\
&&
\left.+\gamma_\pm^2\nu_\pm[\nu_\pm(E_{\hat r\hat r}-E_{\hat \theta\hat \theta})-2H_{\hat r\hat \theta}]
\right\}r_{(1)}^2
\,,\nonumber\\
Z&=&\sqrt{g_{rr}}k_{\rm (lie)}r_{(1)}\nu_{(1)}^{\hat r}-\gamma_\pm^2\nu_\pm^2\nu_{(1)}\alpha_{(1)}+\nu_{(2)}^{\hat r}
\,.
\end{eqnarray}

Similarly, for the 4-momentum $P=mu$, with $u$ given by Eqs. (\ref{polarnuu})--(\ref{polarnuu2}), we have
\begin{eqnarray} 
\nu_u&=&\nu_\pm+\nu_{u(1)}+\nu_{u(2)}\,,\quad
\alpha_u=\frac{\pi}{2}+\alpha_{u(1)}+\alpha_{u(2)}\,,\nonumber\\
\nu_u^{\hat r}&=&\nu^{\hat r}_{u(1)}+\nu^{\hat r}_{u(2)}\,,\quad
\nu_u^{\hat \phi}=\nu_\pm+\nu^{\hat \phi}_{u(1)}+\nu^{\hat \phi}_{u(2)}\,,
\end{eqnarray}
with 
\begin{eqnarray}
\nu^{\hat r}_{u(1)}&=&-\nu_\pm\alpha_{u(1)}\,,\quad
\nu^{\hat r}_{u(2)}=-\nu_{u(1)}\alpha_{u(1)}-\nu_\pm\alpha_{u(2)}\,,\nonumber\\
\nu^{\hat \phi}_{u(1)}&=&\nu_{u(1)}\,,\quad
\nu^{\hat \phi}_{u(2)}=-\frac12\nu_\pm\alpha_{u(1)}^2+\nu_{u(2)}\,.
\end{eqnarray}
Note that both the mass $m=m_0$ of the body and the magnitude of the spin vector $s$ remain constant along the path to that order.

Finally, the spin terms (\ref{fspinframeu}) and (\ref{dspinframeu}) become
\begin{widetext} 
\begin{eqnarray}
F_{\rm (spin)}&=&mM{\hat s}\gamma_\pm^2\left\{
[(1+\nu_\pm^2)H_{\hat r\hat \theta}-\nu_\pm(E_{\hat r\hat r}-E_{\hat \theta\hat \theta})]
-\gamma_\pm^2[(1+\nu_\pm^2)(E_{\hat r\hat r}-E_{\hat \theta\hat \theta})-4\nu_\pm H_{\hat r\hat \theta}]\nu_{(1)}\right.\nonumber\\
&&\left.
+\sqrt{g_{rr}}[(1+\nu_\pm^2)\partial_{\hat r}H_{\hat r\hat \theta}-\nu_\pm\partial_{\hat r}(E_{\hat r\hat r}-E_{\hat \theta\hat \theta})]r_{(1)}
\right\}e_1
+mM{\hat s}\gamma_\pm\left\{
\nu_\pm(2E_{\hat r\hat r}+E_{\hat \theta\hat \theta})-H_{\hat r\hat \theta}
\right\}e_2\,,
\end{eqnarray}
and
\begin{eqnarray}
D_{\rm (spin)}&=&-\omega^0\wedge\left\{
m\gamma_\pm[\nu_\pm(\alpha_{u(2)}-\alpha_{(2)})\omega^{1}
+\gamma_\pm(\nu_{(2)}-\nu_{u(2)})\omega^{2}]\right\}\,,
\end{eqnarray}
whereas the quadrupole terms (\ref{fquadframeu}) and (\ref{dquadframeu})
\begin{eqnarray}
F_{\rm (quad)}&=&F_{\rm (quad)}^1e_1\nonumber\\
&=&
-\frac{2}{3}mM^2\gamma_\pm^2\left\{
\left[(b_1-b_2){\hat X}_{11}+(2b_1-b_3){\hat X}_{22}\right]\nu_\pm^2
+(b_4-b_5)(2{\hat X}_{11}+{\hat X}_{22})\nu_\pm\right.\nonumber\\
&&\left.
+(b_1-b_2){\hat X}_{11}-(b_1+b_2-b_3){\hat X}_{22}
\right\}e_1
\,,
\end{eqnarray}
and
\begin{eqnarray}
D_{\rm (quad)}&=&
-\omega^0\wedge\left\{
-\frac{4}{3}mM^2\gamma^2_\pm(2{\hat X}_{11}+{\hat X}_{22})\left[
(1+\nu_\pm^2)H_{\hat r\hat \theta}-\nu_\pm(E_{\hat r\hat r}-E_{\hat \theta\hat \theta})
\right]\omega^{2}
\right\}\,.
\end{eqnarray}

\end{widetext}

We are interested in solutions which describe deviations from circular geodesic motion, i.e., quasi-circular orbits, due to both the spin-curvature force and the quadrupolar force. 
Hence, we choose initial conditions so that the world line of the extended body has the same starting point as the reference geodesic, i.e., 
\beq
x^{\alpha}_{(1)}(0)=0=x^{\alpha}_{(2)}(0)\,.
\eeq
The two world lines in general have not a common unit tangent vector at $\tau_{\rm (geo)}=0=\tau$; as $\tau$ increases, then, they deviate from each other.
The further requirement that the 4-velocity $U$ be initially tangent to the circular geodesic $U_{\rm (geo)}$ would imply  
\beq
\frac{\rmd x^{\alpha}_{(1)}(0)}{\rmd\tau}=0=\frac{\rmd x^{\alpha}_{(2)}(0)}{\rmd\tau}\,.
\eeq

We will adopt a Keplerian-like parametrization for the orbit \cite{dam-derue1,dam-derue2,wex}, i.e.,
\begin{eqnarray}
\label{quasikeplgen}
\frac{2\pi}{P}(t-t_0)&=& \ell_t -e_t \sin \ell_t\,,\nonumber\\
r&=&a_r(1-e_r \cos \ell_r)\,, \nonumber\\
\theta &=&\frac{\pi}{2}\,,\nonumber\\
\frac{2\pi}{\Phi}(\phi-\phi_0)&=& 2\arctan\left(\sqrt{\frac{1+e_\phi}{1-e_\phi}}\tan\frac{\ell_\phi}{2}\right)\,,
\end{eqnarray}
where $e_t$, $e_r$ and $e_\phi$ are three eccentricities, and $P$ and $\Phi$ denote the periods of $t$ and $\phi$ motions, respectively (with an abuse of notation for $P$, not to be confused with the body's 4-momentum).
The quantities $\ell_t$, $\ell_r$ and $\ell_\phi$ are functions of the proper time parameter $\tau$ on the orbit.
For the circular geodesic (\ref{circgeos}) we have $a_r=r_0$, $e_t=e_r=e_\phi=0$, $P=2\pi\Gamma_\pm$, $\Phi= 2\pi\Omega_\pm$ and $\ell_t=\ell_r=\ell_\phi=\tau$.

The orbital period and the fractional periastron advance \cite{dam88} defined by 
\beq
\label{peradv}
k\equiv\frac{\Phi}{2\pi}-1\,,
\eeq
are directly observable quantities.
We will provide explicit expressions for all orbital elements in terms of gauge-invariant quantities (i.e., total energy and angular momentum) in Section V, where the weak field and slow motion limit of the results of the present section is discussed.

\subsection{Perturbative solution up to the first order}

The first order solution corresponding to the case of a purely spinning particle is given by
\begin{eqnarray} 
\label{solord1}
t_{(1)}&=&{\hat s} 
T_{\hat s}(\sin\Omega_{\rm(ep)}\tau-\Omega_{\rm(ep)}\tau)
+{\hat s}\tilde T_{\hat s}\tau\,,
\nonumber\\
r_{(1)}&=&{\hat s}
R_{\hat s}(\cos\Omega_{\rm(ep)}\tau-1)\,,
\nonumber\\
\phi_{(1)}&=&\bar\zeta_\pm t_{(1)}\,, 
\nonumber\\
\nu_{(1)}&=&{\hat s}
{\mathcal V}^{(\phi)}_{\hat s}(\cos\Omega_{\rm(ep)}\tau-1)
+{\hat s}\tilde{\mathcal V}^{(\phi)}_{\hat s}\,,
\nonumber\\
\alpha_{(1)}&=&
-{\hat s}\frac{{\mathcal V}^{(r)}_{\hat s}}{\nu_\pm}\sin\Omega_{\rm(ep)}\tau\,,
\end{eqnarray}
with
\beq
\nu_{u(1)}=\nu_{(1)}\,,\qquad
\alpha_{u(1)}=\alpha_{(1)}\,.
\eeq
Here
\begin{eqnarray}
R_{\hat s} &=& -\gamma_\pm\frac{\sqrt{\Delta}}{r_0\Omega_{\rm(ep)}}{\mathcal V}^{(r)}_{\hat s}\,,\quad	
T_{\hat s}=\pm2\frac{\gamma_\pm^2\nu_\pm}{N}\frac{\zeta_K}{\Omega_{\rm(ep)}^2}{\mathcal V}^{(r)}_{\hat s}\,,\nonumber\\
{\mathcal V}^{(\phi)}_{\hat s}&=&-\frac{\nu_\pm}{\gamma_\pm}\frac{k_{\rm (lie)}}{\Omega_{\rm(ep)}}{\mathcal V}^{(r)}_{\hat s}\,,\quad
\tilde T_{\hat s}=\frac{\gamma_\pm^3\nu_\pm}{N}\tilde{\mathcal V}^{(\phi)}_{\hat s}\,,\nonumber\\
\tilde{\mathcal V}^{(\phi)}_{\hat s}&=&\pm\frac{\Omega_{\rm(ep)}}{2\gamma_\pm\zeta_K}\left({\mathcal V}^{(r)}_{\hat s}-\tilde{\mathcal V}^{(r)}_{\hat s}\right)\,,\nonumber\\
\tilde{\mathcal V}^{(r)}_{\hat s}&=&-3\frac{M\sqrt{\Delta}}{r_0\Omega_{\rm(ep)}}\frac{\gamma_\pm\zeta_\pm^2}{N^2}\left(\frac{a}{r_0}\mp r_0\zeta_K\right)\,,
\end{eqnarray}
and 
\begin{eqnarray}
\label{omegaep}
\Omega_{\rm(ep)} &=& \zeta_K\left[4-\frac{3\Delta}{r_0^2}\frac{\Omega_\pm^2}{\zeta_K^2}\right]^{1/2}\nonumber\\
&=&|\Omega_\pm|\left[1-\frac{6M}{r_0}-\frac{3a^2}{r_0^2}\pm 8a\zeta_K\right]^{1/2}
\end{eqnarray}
denotes the well known epicyclic frequency governing the radial perturbations of circular geodesics.

Note that one can fix the yet unspecified integration constant ${\mathcal V}^{(r)}_{\hat s}$ in such a way that the 4-velocity $U$ of the spinning particle is tangent to the circular geodesic $U_\pm$ at $\tau=0$, i.e., 
\beq
\nu^{\hat r}_{\hat s}=
{\mathcal V}^{(r)}_{\hat s}\sin(\Omega_{\rm(ep)}\tau) \,,\quad 
\nu^{\hat \phi}_{\hat s}=
{\mathcal V}^{(\phi)}_{\hat s}[\cos(\Omega_{\rm(ep)}\tau)-1]\,,
\eeq
by setting ${\mathcal V}^{(r)}_{\hat s}=\tilde{\mathcal V}^{(r)}_{\hat s}$, implying $\tilde{\mathcal V}^{(\phi)}_{\hat s}=0$.
This is the solution which describes deviations from the equatorial geodesic due to the spin-curvature force. 

The explicit solution of the orbit of the spinning particle can then be also written in the form (\ref{quasikeplgen}) as
\begin{eqnarray}
\label{solorbfinspin}
\frac{2\pi}{P}(t-t_0)&=& \ell -e_t \sin \ell\,,\nonumber\\
r&=&a_r(1-e_r \cos \ell)\,, \nonumber\\
\theta &=&\frac{\pi}{2}\,,\nonumber\\
\frac{2\pi}{\Phi}(\phi-\phi_0)&=& \ell +e_\phi \sin \ell\,,
\end{eqnarray}
where $\ell =\Omega_{\rm(ep)}\tau$ and
\beq
\label{solar}
a_r = r_0(1+e_r)\,,
\eeq
and the eccentricities are given by
\begin{eqnarray}
\label{solecc}
e_t &=& \mp2{\hat s}{\mathcal V}^{(r)}_{\hat s}\frac{\zeta_K}{\Omega_{\rm (ep)}}\frac{N}{\sqrt{\Delta}}\tilde L_\pm
=\mp2{\hat s}{\mathcal V}^{(r)}_{\hat s}\gamma_\pm\nu_\pm\frac{\zeta_K}{\Omega_{\rm (ep)}}\,, \nonumber\\
e_r &=&   {\hat s}{\mathcal V}^{(r)}_{\hat s} \gamma_\pm\frac{\sqrt{\Delta}}{r_0^2\Omega_{\rm (ep)}}\,, \nonumber\\
e_\phi &=&  \pm2{\hat s}{\mathcal V}^{(r)}_{\hat s}\frac{\zeta_K}{\zeta_\pm}\frac{N}{\sqrt{\Delta}\Omega_{\rm (ep)}}\tilde E_\pm
=-\frac{\bar\zeta_\pm}{\zeta_\pm}e_t\,,
\end{eqnarray}
while the periods of $t$ and $\phi$ motions result in
\beq
\label{solper1}
P=2\pi\frac{\Gamma_\pm}{\Omega_{\rm (ep)}}+{\hat s} P_{\hat s}\,, \qquad
\Phi= 2\pi\frac{\Omega_\pm}{\Omega_{\rm (ep)}} +{\hat s} \Phi_{\hat s}\,,
\eeq
with
\begin{eqnarray}
\label{solper2}
P_{\hat s}&=&\pm\pi\frac{\gamma_\pm^2\nu_\pm}{N\zeta_K}\left({\mathcal V}^{(r)}_{\hat s}-\tilde{\mathcal V}^{(r)}_{\hat s}\right)
\mp4\pi\frac{\gamma_\pm^2\nu_\pm}{N}\frac{\zeta_K}{\Omega_{\rm (ep)}^2}{\mathcal V}^{(r)}_{\hat s} \,, \nonumber\\
\Phi_{\hat s}&=& \bar\zeta_\pm P_{\hat s}\,.
\end{eqnarray}
The fractional periastron advance (\ref{peradv}) is thus given by 
\beq
\label{peradvspin}
k=\frac{\Omega_\pm}{\Omega_{\rm (ep)}}-1 +{\hat s} \frac{\Phi_{\hat s}}{2\pi}\,.
\eeq

Finally, the conserved energy and angular momentum (\ref{totalenergy}) per unit mass $\tilde E=E/m$ and $\tilde J=J/m$ are given by
\beq
\label{costmoto1}
\tilde E=\tilde E_\pm+{\hat s} \tilde E_{\hat s}\,, \qquad
\tilde J=\tilde L_\pm+{\hat s}\tilde J_{\hat s}\,,
\eeq
with
\begin{eqnarray}
\label{costmoto2}
\tilde E_{\hat s}&=&\pm\frac{\gamma_\pm\Omega_\pm}{2\zeta_K}\sqrt{\Delta}\Omega_{\rm (ep)}\left({\mathcal V}^{(r)}_{\hat s}-\tilde{\mathcal V}^{(r)}_{\hat s}\right)\nonumber\\
&&
+M\Omega_\pm\left(\frac{M}{r_0}\mp a\zeta_K\right)\,, \nonumber\\
\tilde E_{\hat s}&-&\zeta_\pm\tilde J_{\hat s}= \pm\frac{M\zeta_K}{\Gamma_\pm}\,.
\end{eqnarray}

\subsubsection*{Spinning particles along circular orbits}

The special case in which the orbit of the spinning particle remains circular corresponds to ${\mathcal V}^{(r)}_{\hat s}=0$, that is $e_t=e_r=e_\phi=0$, implying that the 4-velocity is 
\beq
\label{casocircU}
U=\Gamma(\partial_t+\zeta\partial_\phi)\,,
\eeq
with normalization factor
\beq
\Gamma=\Gamma_\pm\left[1\pm \frac32{\hat s}\Omega_\pm^2\nu_\pm\frac{M\sqrt{\Delta}}{r_0\zeta_K}\left(\frac{a}{r_0}\mp r_0\zeta_K\right)\right]\,,
\eeq
and angular velocity  
\beq
\label{zetaspin}
\zeta=\zeta_\pm\left[1\pm \frac32{\hat s}\frac{M\zeta_\pm}{r_0\zeta_K}\left(\frac{a}{r_0}\mp r_0\zeta_K\right)\right]
=\frac{\Phi}{P}\,.
\eeq
The parametric equations of the orbit are given by
\beq
\label{casocircorb}
t=t_0+\Gamma \tau\,,\quad 
r=r_0\,,\quad 
\theta=\frac{\pi}{2}\,,\quad
\phi=\phi_0+\Omega \tau\,,
\eeq
with orbital angular velocity
\beq
\Omega=\Omega_\pm\left[1\pm\frac32{\hat s}r_0^2\Omega_\pm^2\left(1-\frac{2M}{r_0}\pm a\zeta_K\right)\left(\frac{a}{r_0}\mp r_0\zeta_K\right)\right]\,.
\eeq
Note that the fractional periastron advance (\ref{peradv}) does not vanish in the circular case
\beq
\label{peradvcirc}
k=\frac{\Omega}{\Omega_{\rm (ep)}}-1\,.
\eeq
Finally, the conserved quantities are given by Eq. (\ref{costmoto1}) with
\begin{eqnarray}
\label{costmoto2circ}
\tilde E_{\hat s}&=&-\frac{M\Omega_\pm^3}{2\zeta_K^2}\left(\frac{M}{r_0}\mp a\zeta_K\right)\left(1+\frac{3a^2}{r_0^2}\mp 4a\zeta_K\right)\,,\nonumber\\
\tilde J_{\hat s}&=&\mp\frac{M\Omega_\pm^3}{2\zeta_K^3}\left[
\left(\frac{M}{r_0}\mp a\zeta_K\right)\left(1+\frac{3a^2}{r_0^2}\mp 4a\zeta_K\right)\right.\nonumber\\
&&\left.
\times(1\pm a\zeta_K)-2\left(1-\frac{3M}{r_0}\pm 2a\zeta_K\right)^2
\right]\,.
\end{eqnarray}

\subsection{Perturbative solution up to the second order}

The second order solution is given by
\begin{eqnarray} 
\label{solord2}
t_{(2)}&=& D_1\sin\Omega_{\rm(ep)}\tau+D_2\sin2\Omega_{\rm(ep)}\tau+D_3\tau\cos\Omega_{\rm(ep)}\tau\nonumber\\
&&+D_4\tau\,, \nonumber\\
r_{(2)}&=& C_1(\cos\Omega_{\rm(ep)}\tau-1)+C_2(\cos2\Omega_{\rm(ep)}\tau-1)\nonumber\\
&&+C_3\tau\sin\Omega_{\rm(ep)}\tau\,, \nonumber\\
\phi_{(2)}&=& E_1\sin\Omega_{\rm(ep)}\tau+E_2\sin2\Omega_{\rm(ep)}\tau+E_3\tau\cos\Omega_{\rm(ep)}\tau\nonumber\\
&&+E_4\tau\,, \nonumber\\
\nu_{(2)}&=& A_1(\cos\Omega_{\rm(ep)}\tau-1)+A_2(\cos2\Omega_{\rm(ep)}\tau-1)\nonumber\\
&&+A_3\tau\sin\Omega_{\rm(ep)}\tau+A_4\,, \nonumber\\
\alpha_{(2)}&=& B_1\sin\Omega_{\rm(ep)}\tau+B_2\sin2\Omega_{\rm(ep)}\tau+B_3\tau\cos\Omega_{\rm(ep)}\tau\,,\nonumber\\
\end{eqnarray}
where the integration constants are listed in Appendix \ref{const} 
and 
\begin{eqnarray} 
\nu_{u(2)}&=&\nu_{(2)}-\frac{M\Omega_{\rm(ep)}}{\gamma_\pm}\tilde{\mathcal V}^{(r)}_{\hat s}\left[
{\hat s}^2+\frac43(2{\hat X}_{11}+{\hat X}_{22})
\right]\,,\nonumber\\
\alpha_{u(2)}&=&\alpha_{(2)}\,.
\end{eqnarray}
The constant $A_4$ remains undetermined.
Choosing $A_4=0$ corresponds to require that the 4-velocity $U$ of the extended body be tangent to the circular geodesic $U_\pm$ at $\tau=0$. 

The explicit solution of the orbit of the extended body can then be also written in the form (\ref{quasikeplgen}) as
\begin{eqnarray}
\label{solorbfinquad}
\frac{2\pi}{P}(t-t_0)&=& \ell_t -e_t \sin \ell_t\,,\nonumber\\
r&=&a_r(1-e_r \cos \ell_r)\,, \nonumber\\
\theta &=&\frac{\pi}{2}\,,\nonumber\\
\frac{2\pi}{\Phi}(\phi-\phi_0)&=& \ell_\phi +e_\phi \sin \ell_\phi+\frac{e_\phi^2}{4}\sin2\ell_\phi\,,
\end{eqnarray}
where $\ell_t$, $\ell_r$ and $\ell_\phi$ are function of $\ell =\Omega_{\rm(ep)}\tau$ and are given by
\begin{eqnarray}
\ell_t&=&\ell+\frac{{\hat s}^2}{\Gamma_\pm}[D_{2{\hat s}}\Omega_{\rm (ep)}\sin2\ell+D_{3{\hat s}}\ell\cos \ell]\,,\nonumber\\
\ell_r&=&\ell+\frac{{\hat s}}{R_{\hat s}\Omega_{\rm (ep)}}[2C_{2{\hat s}}\Omega_{\rm (ep)}\sin\ell-C_{3{\hat s}}\ell]\,,\nonumber\\
\ell_\phi&=&\ell+\frac{{\hat s}^2}{\Omega_\pm}\left[
\left(E_{2{\hat s}}\Omega_{\rm (ep)}
-\frac{1}{4}\Omega_{\rm (ep)}^2\bar\zeta_\pm^2T_{\hat s}^2\right.\right.\nonumber\\
&&\left.\left.
\times(T_{\hat s}\Omega_{\rm (ep)}-\tilde T_{\hat s})
\right)\sin2\ell+E_{3{\hat s}}\ell\cos \ell\right]\,.
\end{eqnarray}
The semi-major axis and the eccentricities turn out to be
\beq
\label{solarquad}
a_r = r_0-{\hat s}R_{\hat s}-C_1\,,
\eeq
and 
\begin{eqnarray}
\label{soleccquad}
e_t &=&-\frac{\Omega_{\rm (ep)}}{\Gamma_\pm}\left[
{\hat s}T_{\hat s}+D_1+{\hat s}^2\frac{T_{\hat s}}{\Gamma_\pm}(T_{\hat s}\Omega_{\rm (ep)}-\tilde T_{\hat s})
\right]\,, \nonumber\\
e_r &=&-{\hat s}\frac{R_{\hat s}}{r_0}-\frac{C_1}{r_0}-{\hat s}^2\frac{R_{\hat s}^2}{r_0^2}\,, \\
e_\phi &=&\frac{\Omega_{\rm (ep)}}{\Omega_\pm}\left[
\bar\zeta_\pm{\hat s}T_{\hat s}+E_1+{\hat s}^2\frac{T_{\hat s}}{\Omega_\pm}\bar\zeta_\pm^2(T_{\hat s}\Omega_{\rm (ep)}-\tilde T_{\hat s})
\right]\,,\nonumber
\end{eqnarray}
respectively, whereas the periods of $t$ and $\phi$ motions read
\begin{eqnarray}
\label{solper1quad}
P&=&2\pi\frac{\Gamma_\pm}{\Omega_{\rm (ep)}}+{\hat s} P_{\hat s}+\frac{2\pi}{\Omega_{\rm (ep)}}D_4\,, \nonumber\\
\Phi&=& 2\pi\frac{\Omega_\pm}{\Omega_{\rm (ep)}} +{\hat s} \Phi_{\hat s}+\frac{2\pi}{\Omega_{\rm (ep)}}E_4\,,
\end{eqnarray}
where $P_{\hat s}$ and $\Phi_{\hat s}$ are given by Eq. (\ref{solper2}).
The fractional periastron advance (\ref{peradv}) is then 
\beq
\label{peradvquad}
k=\frac{\Omega_\pm}{\Omega_{\rm (ep)}}-1 +{\hat s} \frac{\Phi_{\hat s}}{2\pi}+\frac{E_4}{\Omega_{\rm (ep)}}\,.
\eeq

Finally, the conserved energy and angular momentum per unit mass (\ref{totalenergy}) turn out to be
\beq
\label{costmoto1quad}
\tilde E=\tilde E_\pm+{\hat s}\tilde E_{\hat s}+\tilde E_{(2)}\,, \qquad
\tilde J=\tilde L_\pm+{\hat s}\tilde J_{\hat s}+\tilde J_{(2)}\,,
\eeq
with $\tilde E_{\hat s}$ and $\tilde J_{\hat s}$ as in Eq. (\ref{costmoto2}) and 
\begin{eqnarray}
\label{costmoto2quad}
\tilde J_{(2)}&=&\sqrt{g_{\phi\phi}}\gamma_\pm^2\left\{
\gamma_\pm A_4-\frac{4}{3}M\Omega_{\rm (ep)}\tilde{\mathcal V}^{(r)}_{\hat s}(2{\hat X}_{11}+{\hat X}_{22})\right.\nonumber\\
&&\left.
+{\hat s}^2\left[
-\frac{2}{\gamma_\pm\nu_\pm}({\mathcal V}^{(r)}_{\hat s})^2+\frac32\gamma_\pm^3\nu_\pm(\tilde{\mathcal V}^{(\phi)}_{\hat s})^2\right.\right.\nonumber\\
&&\left.\left.
\pm3\gamma_\pm M\zeta_K\tilde{\mathcal V}^{(\phi)}_{\hat s}-M\Omega_{\rm (ep)}{\mathcal V}^{(r)}_{\hat s}
\right]
\right\}\,,\nonumber\\
\tilde E_{(2)}&-&\zeta_\pm\tilde J_{(2)}=2N\gamma_\pm{\hat s}^2\left[
({\mathcal V}^{(r)}_{\hat s})^2+\frac{\gamma_\pm^2}{4}(\tilde{\mathcal V}^{(\phi)}_{\hat s})^2
\right]\,.
\end{eqnarray}

\subsubsection*{Quadrupolar bodies along circular orbits}

The special case in which the orbit of the extended body remains circular corresponds to ${\mathcal V}^{(r)}_{\hat s}=0$ and $C_1=C_2=C_3=0$, implying that all second order coefficients are zero except for $A_4$, $D_4$ and $E_4$, which become
\begin{eqnarray}
A_4&=&\mp\frac{1}{2m\gamma_\pm^2\zeta_K}\left(F^1_{\rm (quad)}\pm\zeta_K{\mathcal E}(u)_{\rm (quad)}^2\right)\nonumber\\
&&
+\frac{{\hat s}^2}{2M\zeta_K}\left[
M(k_{\rm (lie)}\mp4\gamma_\pm^2\nu_\pm\zeta_K)\tilde{\mathcal V}^{(\phi)}_{\hat s}\right.\nonumber\\
&&\left.
-M^2(7\zeta_K^2-2\Omega_{\rm(ep)}^2)
\right]\tilde{\mathcal V}^{(\phi)}_{\hat s}
\,,\nonumber\\
D_4&=&
\Gamma_\pm\gamma_\pm^2\left[\nu_\pm A_4+\frac12\gamma_\pm^2(1+2\nu_\pm^2){\hat s}^2(\tilde{\mathcal V}^{(\phi)}_{\hat s})^2\right]
\,,\nonumber\\
E_4&=&
\bar\zeta_\pm D_4
-\frac{\gamma_\pm^3}{2\nu_\pm\sqrt{g_{\phi\phi}}}{\hat s}^2(\tilde{\mathcal V}^{(\phi)}_{\hat s})^2\,.
\end{eqnarray}

The 4-velocity is given by Eq. (\ref{casocircU}) with normalization factor
\beq
\Gamma=\Gamma_\pm\left[
1+\gamma_\pm^2\nu_\pm{\hat s}\tilde{\mathcal V}^{(\phi)}_{\hat s}+\frac{D_4}{\Gamma_\pm}
\right]\,,
\eeq
and angular velocity  
\beq
\label{zetaquad}
\zeta=\zeta_\pm\left[1+\frac{N}{\zeta_\pm\sqrt{g_{\phi\phi}}}\left({\hat s}\tilde{\mathcal V}^{(\phi)}_{\hat s}+A_4\right)\right]
=\frac{\Phi}{P}\,.
\eeq
The parametric equations of the orbit are given by Eq. (\ref{casocircorb}) with orbital angular velocity
\beq
\Omega=\Omega_\pm\left[1+\frac{\bar\zeta_\pm}{\zeta_\pm}\gamma_\pm^2\nu_\pm{\hat s}\tilde{\mathcal V}^{(\phi)}_{\hat s}+\frac{E_4}{\Omega_\pm}\right]
=\frac{\Phi}{2\pi}\Omega_{\rm(ep)}\,.
\eeq
The fractional periastron advance is still given by Eq. (\ref{peradvcirc}).

Finally, the conserved quantities are given by Eq. (\ref{costmoto1quad}) with $\tilde E_{\hat s}$ and $\tilde J_{\hat s}$ as in Eq. (\ref{costmoto2circ}) and 
\begin{eqnarray}
\tilde E_{(2)}&=&\zeta_\pm\tilde J_{(2)}+\frac12N\gamma_\pm^3{\hat s}^2(\tilde{\mathcal V}^{(\phi)}_{\hat s})^2\,,\nonumber\\
\tilde J_{(2)}&=&\sqrt{g_{\phi\phi}}\gamma_\pm^2\left[
\gamma_\pm A_4-\frac{4}{3}M\Omega_{\rm (ep)}\tilde{\mathcal V}^{(r)}_{\hat s}(2{\hat X}_{11}+{\hat X}_{22})\right.\nonumber\\
&&\left.
+\frac32\gamma_\pm^2{\hat s}^2\left(
\gamma_\pm\nu_\pm\tilde{\mathcal V}^{(\phi)}_{\hat s}\pm2M\zeta_K
\right)\tilde{\mathcal V}^{(\phi)}_{\hat s}
\right]\,.
\end{eqnarray}

\subsection{Comparison with numerically-integrated orbits}

As we have seen, the first order solution (\ref{solorbfinspin}) is characterized by an oscillatory behavior of the radial component about the geodesic orbit in a circular ring either inside or outside the geodesic radius depending on the relative sign of the vertical component of the spin and the orbital velocity.  
The azimuthal motion also oscillates around the geodesic value with the same frequency characterizing the radial motion, apart from a secular drift which occurs at slightly different speeds for the inner and outer radial oscillations (see also Ref. \cite{spin_dev_schw}).

\begin{figure*} 
\typeout{*** EPS figure 1}
\begin{center}
$\begin{array}{cc}
\includegraphics[scale=0.4]{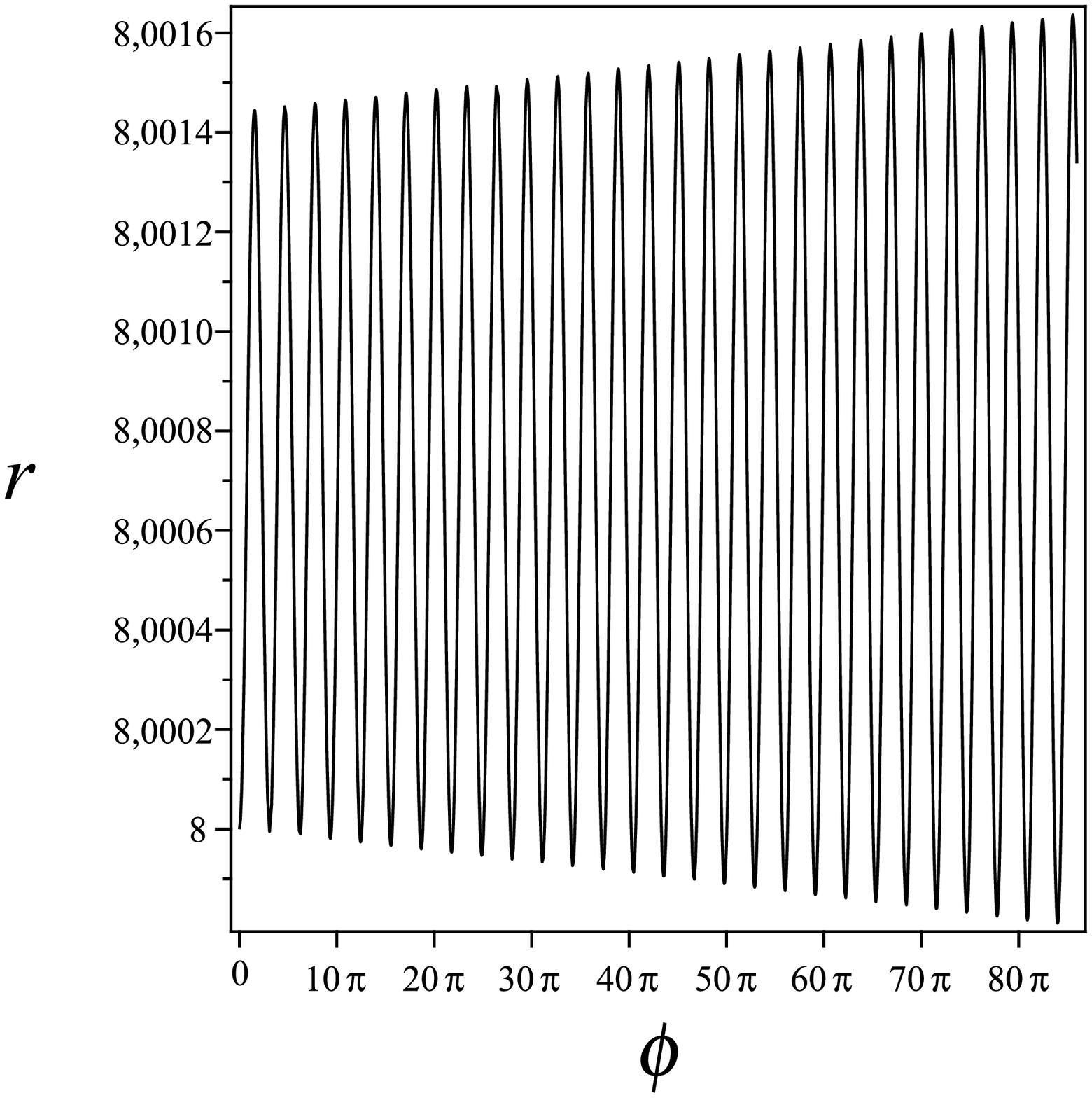}&\quad
\includegraphics[scale=0.4]{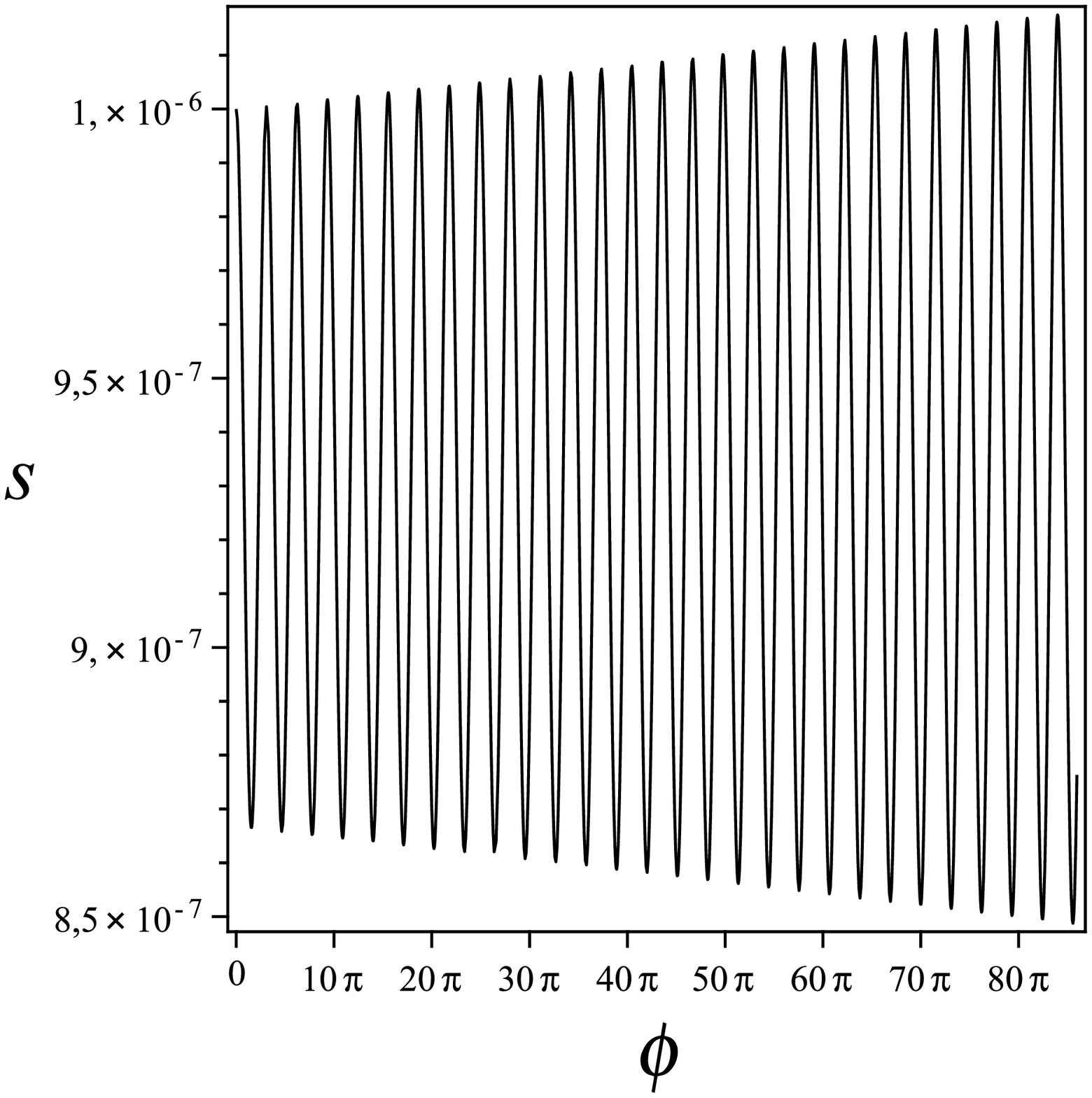}\\[.4cm]
\quad\mbox{(a)}\quad &\quad \mbox{(b)}
\end{array}$\\
\end{center}
\caption{The behavior of the radial coordinate (in units of $M$) is shown in panel (a) as a function of the azimuthal coordinate by solving the whole set of MPD equations numerically with the following choice of parameters and initial conditions: $a/M=0.5$, $r_0/M=8$, $X(u)_{11}/(m_0M^2)=0.001$, $X(u)_{22}/(m_0M^2)=-0.003$ and $r(0)=r_0$, $\phi(0)=0$, $\alpha_u(0)=\pi/2$, $\nu_u(0)=\nu_+\approx0.38$, $m(0)=m_0$, $s(0)=10^{-6}$ (in units of $m_0M$).
Panel (b) shows instead the corresponding behavior of the spin magnitude.
}
\label{fig:1}
\end{figure*}

The second order solutions (\ref{solorbfinquad}) are still oscillatory as those of first order, but with two different frequencies, the epicyclic one and twice it.
Furthermore, the second order quantities all contain secular terms which increase with proper time and cause the widening after each revolution of the region wherein the motion is confined to.
This result is confirmed by solving numerically the full set of nonlinear equations (see Appendix C) for very small values of both spin and quadrupole parameters.  
Figs. \ref{fig:1} (a) and (b) show the behaviors of the radial coordinate and of the spin invariant respectively as functions of the azimuthal coordinate in the case of very small values of the spin parameter as well as of the non-vanishing constant frame components $X(u)_{11}$ and $X(u)_{22}$ of the quadrupole tensor.
The initial conditions have been chosen so that the tangent vector $U$ to the world line of the extended body is initially aligned with a stable  equatorial circular geodesic at a given value of the radial coordinate and the body has initially negligible spin.
We see that the motion is confined inside a band close to the circular geodesic whose thickness slightly increases after each revolution.
This feature is definitely new with respect to the case of a purely spinning particle, where the orbit oscillates filling a circular corona of fixed width \cite{mashsingh,spin_dev_schw}.
Nevertheless, it was already found to occur in the case of a quadrupolar body moving in a Schwarzschild background \cite{quadrupschw}. 
The occurrence of such a secular increase of the bandwidth for small values of the quadrupole is actually a higher order effect only.
In fact, if the body is initially endowed with spin the oscillation amplitude is almost fixed by the value of the spin parameter.


\section{Weak field and slow motion}

Let us consider now the weak field limit of the above analysis.
Introduce the dimensionless quantities $u_0=M/r_0$ and $\hat a=a/M$.
In the case in which the 4-velocity $U$ of the extended body is initially tangent to the reference circular geodesic $U_{\rm(geo)}$, the expansion of the conserved energy and angular momentum per unit mass of the body reads  

\begin{eqnarray}
-2{\widehat E}&\equiv&-2(\tilde E-1)\nonumber\\
&=&-2{\widehat E}_{\rm(geo)}
+2{\hat s}u_0^{5/2}\left(1-{\hat a}u_0^{1/2}+\frac32u_0\right)\nonumber\\
&&
-\left[5{\hat a}{\hat s}-6{\hat s}^2-8(2{\hat X}_{11}+{\hat X}_{22})\right]u_0^4
+O(u_0^{9/2})
\,,\nonumber\\
\end{eqnarray}
and
\begin{eqnarray}
{\widehat J}+\hat s&\equiv&\frac{\tilde J}{M}+\hat s\nonumber\\
&=&{\widehat J}_{\rm(geo)}+\hat s\left(1+\frac12u_0-\frac38u_0^2\right)
\nonumber\\
&&
+\left[-21{\hat s}^2-4(2{\hat X}_{11}+{\hat X}_{22})\right]u_0^{5/2}
+O(u_0^3)\,, \nonumber\\
\end{eqnarray}
where 
\begin{eqnarray}
-2{\widehat E}_{\rm(geo)}&=&u_0\left(1-\frac34u_0-\frac{27}{8}u_0^2-\frac{675}{64}u_0^3\right)\nonumber\\
&&
+2{\hat a}u_0^{5/2}\left(1+\frac92u_0\right)-{\hat a}^2u_0^3\left(1+\frac{15}{2}u_0\right)\nonumber\\
&&
+O(u_0^{5})
\,,\nonumber\\
{\widehat J}_{\rm(geo)}&=&u_0^{-1/2}\left(1+\frac32u_0+\frac{27}{8}u_0^2+\frac{135}{16}u_0^3\right)\nonumber\\
&&
-3{\hat a}u_0\left(1+\frac{5}{2}u_0\right)+{\hat a}^2u_0^{3/2}\left(1+5u_0\right)\nonumber\\
&&
+O(u_0^3)\,,
\end{eqnarray}
denote the corresponding geodesic quantities, and terms higher than the second in the background rotation parameter have been neglected.
Note that we have selected corotating circular geodesics by choosing the plus sign in Eq. (\ref{Ugeodef}) and related quantities, the counter-rotating case being simply obtained by $a\to-a$.   

We list below the orbital elements (\ref{solarquad})--(\ref{solper1quad}) expressed in terms of the gauge invariant quantities ${\widehat E}$ and ${\widehat J}$.
Note that the latter are related by

\begin{widetext} 

\begin{eqnarray}
{\widehat J}&=&(-2{\widehat E})^{-1/2}\left\{
1+(-2{\widehat E})\left[
\frac98+\frac{243}{128}(-2{\widehat E})+\frac{5373}{1024}(-2{\widehat E})^2+\frac{1173123}{32768}(-2{\widehat E})^3
\right]\right.\nonumber\\
&&\left.
+(-2{\widehat E})^2\left[\frac{\hat a}{2}(\hat a-2\hat s)
+\frac32\left(\frac{23}{8}{\hat a}^2-\frac{15}{4}{\hat a}{\hat s}-15{\hat s}^2\right)(-2{\widehat E})\right.\right.\nonumber\\
&&\left.\left.
+\left(\frac{13923}{256}{\hat a}^2-\frac{4275}{128}{\hat a}{\hat s}-\frac{2703}{16}{\hat s}^2-10(2{\hat X}_{11}+{\hat X}_{22})\right)(-2{\widehat E})^2
\right]\right.\nonumber\\
&&\left.
+(-2{\widehat E})^{1/2}\left[
-{\hat s}+\frac12(3\hat s-4\hat a)(-2{\widehat E})+\frac92(\hat s-2\hat a)(-2{\widehat E})^2\left(1+\frac{9}{4}(-2{\widehat E})\right)
\right]
\right\}
+O[(-2{\widehat E})^{7/2}]\,.
\end{eqnarray}

We find 
\begin{eqnarray}
\frac{a_r}{M}&=&\frac1{u_0}+3\sqrt{u_0}{\hat s}(1+4u_0+24u_0^2+144u_0^3)-3({\hat a}{\hat s}-2{\hat X}_{11})u_0+(15{\hat s}^2-36{\hat a}{\hat s}+10{\hat X}_{11}+2{\hat X}_{22})u_0^2
+O(u_0^3)
\nonumber\\
&=&(-2{\widehat E})^{-1}\left[
1-\frac34(-2{\widehat E})+(2{\hat a}+5{\hat s})(-2{\widehat E})^{3/2}+\left(-\frac{63}{16}-5{\hat a}{\hat s}-{\hat a}^2+6{\hat X}_{11}\right)(-2{\widehat E})^2\right.\nonumber\\
&&\left.
+\left(\frac{51}{4}{\hat a}+\frac{159}{8}{\hat s}\right)(-2{\widehat E})^{5/2}
+\left(-\frac{1215}{64}-\frac{251}{4}{\hat a}{\hat s}-\frac{63}{4}{\hat a}^2+12{\hat s}^2+\frac{61}{2}{\hat X}_{11}+10{\hat X}_{22}\right)(-2{\widehat E})^3\right]\nonumber\\
&&
+O[(-2{\widehat E})^{5/2}]
\,,\nonumber\\
e_t&=&-6{\hat s}u_0^{5/2}(1+6u_0+36u_0^2)+6({\hat a}{\hat s}-2{\hat X}_{11})u_0^3
+(33{\hat s}^2+102{\hat a}{\hat s}-44{\hat X}_{11}-4{\hat X}_{22})u_0^4
+O(u_0^5)
\nonumber\\
&=&(-2{\widehat E})^{5/2}\left[-6 {\hat s} +6({\hat a} {\hat s}-2{\hat X}_{11})(-2{\widehat E})^{1/2}
-\frac{189}{4}{\hat s}(-2{\widehat E})
+\left(\frac{291}{2}{\hat a}{\hat s}+63{\hat s}^2-71{\hat X}_{11}-4{\hat X}_{22}\right)(-2{\widehat E})^{3/2}
\right]\nonumber\\
&&
+O[(-2{\widehat E})^{9/2}]
\,,\nonumber\\
e_r&=&3{\hat s}u_0^{3/2}(1+4u_0+24u_0^2)-3({\hat a}{\hat s}-2{\hat X}_{11})u_0^2
+2(3{\hat s}^2-18{\hat a}{\hat s}+5{\hat X}_{11}+{\hat X}_{22})u_0^3
+O(u_0^4)\nonumber\\
&=&(-2{\widehat E})^{3/2}\left[3 {\hat s} -3({\hat a} {\hat s}-2{\hat X}_{11})(-2{\widehat E})^{1/2}
+\frac{123}{8}{\hat s}(-2{\widehat E})
+\left(-\frac{99}{2}{\hat a}{\hat s}-3{\hat s}^2+19{\hat X}_{11}+2{\hat X}_{22}\right)(-2{\widehat E})^{3/2}
\right]\nonumber\\
&&
+O[(-2{\widehat E})^{7/2}]
\,,\nonumber\\
e_\phi&=&6{\hat s}u_0^{3/2}(1+4u_0+24u_0^2)-6({\hat a}{\hat s}-2{\hat X}_{11})u_0^2
-2(3{\hat s}^2+33{\hat a}{\hat s}-10{\hat X}_{11}-2{\hat X}_{22})u_0^3
+O(u_0^4)\nonumber\\
&=&(-2{\widehat E})^{3/2}\left[6 {\hat s} -6({\hat a} {\hat s}-2{\hat X}_{11})(-2{\widehat E})^{1/2}
+\frac{123}{4}{\hat s}(-2{\widehat E})
+\left(-93{\hat a}{\hat s}-24{\hat s}^2+38{\hat X}_{11}+4{\hat X}_{22}\right)(-2{\widehat E})^{3/2}
\right]\nonumber\\
&&
+O[(-2{\widehat E})^{7/2}]
\,.
\end{eqnarray}

The three eccentricities are related by
\begin{eqnarray}
\frac{e_t}{e_r}&=&-2(-2{\widehat E})\left[
1+\frac{11}{4}(-2{\widehat E})
-\left(5{\hat a}+\frac{19}{2}{\hat s}\right)(-2{\widehat E})^{3/2}
+\left(\frac{23}{2}+{\hat a}^2+2{\hat a}{\hat s}\right)(-2{\widehat E})^2
-\left(\frac{283}{8}{\hat a}+\frac{1277}{16}{\hat s}\right)(-2{\widehat E})^{5/2}\right.\nonumber\\
&&\left.
+\left(\frac{983}{16}+93{\hat a}{\hat s}+\frac{161}{4}{\hat a}^2-16{\hat X}_{11}-8{\hat X}_{22}\right)(-2{\widehat E})^{3}
\right]
+O[(-2{\widehat E})^{9/2}]
\,,\nonumber\\
\frac{e_\phi}{e_r}&=&2\left[
1-{\hat s}^2(-2{\widehat E})+({\hat a}-3{\hat s})(-2{\widehat E})^{3/2}+\frac{33}{4}{\hat s}^2(-2{\widehat E})^2+\left(\frac{25}{8}{\hat a}-\frac{147}{8}{\hat s}\right)(-2{\widehat E})^{5/2}
-(3{\hat a}^2-6{\hat a}{\hat s}+39{\hat s}^2)(-2{\widehat E})^{3}\right.\nonumber\\
&&\left.
+\left(\frac{321563}{128}{\hat a}-\frac{22401}{128}{\hat s}\right)(-2{\widehat E})^{7/2}
-\left(\frac{396819}{128}-\frac{247}{2}{\hat a}{\hat s}+\frac{139}{4}{\hat a}^2+\frac{609}{4}{\hat s}^2\right)(-2{\widehat E})^{4}\right.\nonumber\\
&&\left.
+\left(\frac{3005685}{1024}{\hat a}+\frac{6165801}{1024}{\hat s}\right)(-2{\widehat E})^{9/2}\right.\nonumber\\
&&\left.
-\left(\frac{942625431}{131072}-\frac{5297673}{512}{\hat a}{\hat s}+\frac{5632025}{1024}{\hat a}^2+\frac{390501}{128}{\hat s}^2-135{\hat X}_{11}-\frac{135}{2}{\hat X}_{22}\right)(-2{\widehat E})^{5}
\right]
+O[(-2{\widehat E})^{11/2}]
\,.
\end{eqnarray}

Finally, the periods of $t$ and $\phi$ motions are given by
\begin{eqnarray}
\frac1M\frac{P}{2\pi}&=&\frac1{u_0^{3/2}}\left(1+3u_0+\frac{27}{2}u_0^2+\frac{135}{2}u_0^3+\frac{33939}{128}u_0^4\right)
-3{\hat a}+\frac32{\hat a}^2\sqrt{u_0}-3(11{\hat a}+2{\hat s})u_0
+\left(\frac{67}{2}{\hat a}^2+6{\hat a}{\hat s}-12{\hat X}_{11}\right)u_0^{3/2}\nonumber\\
&&
-\left(\frac{513}{2}{\hat a}+54{\hat s}\right)u_0^2
+\left(\frac{4575}{16}{\hat a}^2+138{\hat a}{\hat s}+33{\hat s}^2-80{\hat X}_{11}-4{\hat X}_{22}\right)u_0^{5/2}
+O(u_0^3)\nonumber\\
&=&(-2{\widehat E})^{-3/2}\left[
1+\frac{15}{8}(-2{\widehat E})+3{\hat s}(-2{\widehat E})^{3/2}
+\left(\frac{855}{128}-3{\hat a}{\hat s}\right)(-2{\widehat E})^2
-6(2{\hat a}-{\hat s})(-2{\widehat E})^{5/2}\right.\nonumber\\
&&\left.
+\left(\frac{41175}{1024}+11{\hat a}^2-\frac{201}{8}{\hat a}{\hat s}+\frac32{\hat s}^2+12{\hat X}_{11}+12{\hat X}_{22}\right)(-2{\widehat E})^3
\right]
+O[(-2{\widehat E})^{7/2}]
\,,\nonumber\\
\frac{\Phi}{2\pi}&=&1+3u_0+\frac{27}{2}u_0^2+\frac{135}{2}u_0^3-2(2{\hat a}+3{\hat s})u_0^{3/2}+\left(\frac32{\hat a}^2+6{\hat a}{\hat s}-12{\hat X}_{11}\right)u_0^2
-6(6{\hat a}+7{\hat s})u_0^{5/2}\nonumber\\
&&
+\left(\frac{75}{2}{\hat a}^2+108{\hat a}{\hat s}+\frac{75}{2}{\hat s}^2-56{\hat X}_{11}-4{\hat X}_{22}\right)u_0^3
-27(10{\hat a}+11{\hat s})u_0^{7/2}
+O(u_0^4)\nonumber\\
&=&
1+3(-2{\widehat E})-2(2{\hat a}+3{\hat s})(-2{\widehat E})^{3/2}
+\left(\frac{63}{4}+6{\hat a}{\hat s}+\frac32{\hat a}^2-12{\hat X}_{11}\right)(-2{\widehat E})^2
-\left(\frac{93}{2}{\hat a}+\frac{219}{4}{\hat s}\right)(-2{\widehat E})^{5/2}\nonumber\\
&&
+\left(\frac{405}{4}+\frac{219}{4}{\hat a}^2+153{\hat a}{\hat s}+\frac{111}{2}{\hat s}^2-74{\hat X}_{11}-4{\hat X}_{22}\right)(-2{\widehat E})^3
+O[(-2{\widehat E})^{7/2}]
\,.
\end{eqnarray}

\end{widetext}

The corresponding expression for the fractional periastron advance (\ref{peradv}) then immediately follows.

\subsection{Circular orbits}

In the case of circular orbits we get  
\begin{eqnarray}
-2{\widehat E}&=&-2{\widehat E}_{\rm(geo)}
-{\hat s}u_0^{5/2}\left(1+\frac92u_0\right)+\left({\hat a}{\hat s}+6{\hat X}_{11}\right)u_0^3\nonumber\\
&&
+\left[\frac{23}{2}{\hat a}{\hat s}+\frac32{\hat s}^2+17{\hat X}_{11}+10{\hat X}_{22}\right]u_0^4\nonumber\\
&&
+O(u_0^{9/2})
\,,\nonumber\\
{\widehat J}+\hat s&=&{\widehat J}_{\rm(geo)}
+2{\hat s}u_0\left(1+\frac{27}{16}u_0\right)-\frac32\left({\hat a}{\hat s}+2{\hat X}_{11}\right)u_0^{3/2}\nonumber\\
&&
-\frac12\left[\frac{27}{2}{\hat a}{\hat s}+\frac{15}{4}{\hat s}^2+17{\hat X}_{11}+10{\hat X}_{22}\right]u_0^{5/2}\nonumber\\
&&
+O(u_0^3)
\,.
\end{eqnarray}
They are related by

\begin{widetext} 

\begin{eqnarray}
{\widehat J}&=&(-2{\widehat E})^{-1/2}\left\{
1+(-2{\widehat E})\left[
\frac98+\frac{243}{128}(-2{\widehat E})+\frac{5373}{1024}(-2{\widehat E})^2
\right]
+\frac{\hat a}{2}(-2{\widehat E})^2\left[\hat a-2\hat s+\frac38(23\hat a-30\hat s)(-2{\widehat E})\right]\right.\nonumber\\
&&\left.
+(-2{\widehat E})^{1/2}\left[
-{\hat s}+\frac12(3\hat s-4\hat a)(-2{\widehat E})+\frac92(\hat s-2\hat a)(-2{\widehat E})^2\left(1+\frac{9}{4}(-2{\widehat E})\right)
\right]
\right\}
\,,
\end{eqnarray}
where the expansion has been truncated at the same order of approximation as above.
The periods of $t$ and $\phi$ motions are given by

\begin{eqnarray}
\frac{1}{M}\frac{P}{2\pi}&=&(-2{\widehat E})^{-3/2}\left[
1+\frac{15}{8}(-2{\widehat E})+\frac{855}{128}(-2{\widehat E})^2+\frac{41175}{1024}(-2{\widehat E})^3+\frac{6477003}{32768}(-2{\widehat E})^4
\right]\nonumber\\
&&
-\left[\frac32{\hat s}+12({\hat a}+{\hat s})(-2{\widehat E})+\frac14\left(\frac{531}{2}{\hat a}+297{\hat s}\right)(-2{\widehat E})^2
+\left(\frac{1809}{4}{\hat a}+567{\hat s}\right)(-2{\widehat E})^3
\right]\nonumber\\
&&
+(-2{\widehat E})^{1/2}\left\{
\frac32({\hat a}{\hat s}+6{\hat X}_{11})
+(-2{\widehat E})\left[
-\frac{25}{4}{\hat a}^2+\frac{489}{16}{\hat a}{\hat s}+\frac{3}{8}{\hat s}^2+\frac{195}{8}{\hat X}_{11}+3{\hat X}_{22}
\right]
\right\}
\,,\nonumber\\
\frac{\Phi}{2\pi}&=&1+(-2{\widehat E})\left[
3+\frac{63}{4}(-2{\widehat E})+\frac{405}{4}(-2{\widehat E})^2
\right]
-\frac12(-2{\widehat E})^{3/2}\left[
8{\hat a}+3{\hat s}+(-2{\widehat E})\left(93{\hat a}+\frac{75}{8}{\hat s}\right)
\right]\nonumber\\
&&
+(-2{\widehat E})^2\left\{
\frac32{\hat a}({\hat a}+{\hat s})-3{\hat X}_{11}
+(-2{\widehat E})\left[
\frac{219}{2}{\hat a}^2+\frac{45}{4}{\hat a}{\hat s}+\frac38{\hat s}^2-\frac{55}{2}{\hat X}_{11}-{\hat X}_{22}
\right]\right\}\,,
\end{eqnarray}
so that the angular velocity turns out to be
\begin{eqnarray}
M\zeta&=&M\frac{\Phi}{P}
=(-2{\widehat E})^{3/2}\left[
1+\frac98(-2{\widehat E})-4{\hat a}(-2{\widehat E})^{3/2}+\left(\frac{891}{128}+\frac32{\hat a}
^2-12{\hat X}_{11}\right)(-2{\widehat E})^2
-9(3{\hat a}-{\hat s})(-2{\widehat E})^{5/2}\right.\nonumber\\
&&\left.
+\left(
\frac{41445}{1024}+\frac{655}{16}{\hat a}^2-27{\hat a}{\hat s}-\frac{127}{2}{\hat X}_{11}-16{\hat X}_{22}
\right)(-2{\widehat E})^3
\right]\,,
\end{eqnarray}

\end{widetext}
which is a gauge-invariant quantity as well.

\section{Concluding remarks}

We have investigated the dynamics of extended bodies endowed with intrinsic spin and quadrupole moment in the Kerr spacetime according to the MPD model, extending previous works \cite{spin_dev_schw,mashsingh,quadrupschw}.
As it is well known, the quadrupole tensor has in general 20 independent components and obeys all the algebraic symmetries of the Riemann tensor. 
We have shown that, since this tensor enters the MPD equations only through certain combinations (contractions), the number of effective components actually reduces from 20 to 10, as expected from the standard post-Newtonian formulation of motion of many-body systems. This fact relating the MPD model with approximate treatments was never highlighted in the literature.

We have assumed here that the motion of the extended body be confined on the equatorial plane of a Kerr background, and its spin vector orthogonal to it.
This request is consistent with an even lesser number of effective components of the quadrupole tensor, which reduces from 10 to 5 the number of degrees of freedom.
Moreover, our analysis only concerns  \lq\lq quasi-rigid'' bodies in the sense of Ehlers and Rudolph \cite{ehlers77}, i.e., with quadrupole tensor having constant components with respect to the frame adapted to the body's generalized 4-momentum, as the most natural and simplifying choice. 
Furthermore, we have fully explored both analytically and numerically the case in which the quadrupole tensor is completely specified by two independent components.
Imposing such a condition is not so restrictive, as it does not affect the essential features of the underlying physics.
It corresponds to have the quadrupole tensor purely electric  in the rest frame of the body, and also diagonal. A similar choice was done for example in Ref. 
\cite{hinderer}, which formally implies in our notation fixing the two non-vanishing quadrupole components equal among them (thus further reducing the quadrupole degrees of freedom to a single independent component) and proportional to the square of the spin of the body (so that the quadrupole tensor is only  spin-induced). 

The presence of the quadrupole significantly changes the features of the motion with respect to the case of a purely spinning body.
In fact, both the mass and the magnitude of the spin vector are no longer constant along the path, as a general result.
Furthermore, the quadrupolar structure of the body is responsible for the onset of spin angular momentum, if the body is initially not spinning.
This is evident either from the numerical integration of the orbits and the analytic perturbative solution of the MPD equations which we have obtained in the limit of small spin and quadrupole parameters, i.e., when the characteristic length scales associated with the spin and quadrupole are taken to be very small with respect to the background curvature characteristic length, which is the limit of validity of the MPD model.  

In particular, we have considered the case in which the world line of the extended body stems from the same spacetime point as a stable equatorial circular geodesic, then distinguishing among the cases in which the associated tangent vectors are initially aligned or not. 
We have shown that for a special choice of integration constants the extended body is allowed to move along a spatially circular orbit. 
However, in general, the trajectory oscillates, filling a nearly circular corona of fixed width (depending on the chosen values of the spin and quadrupole parameters) around the geodesic path, similarly to the case of a purely spinning particle.
It is the presence of secular terms in the solution for the center of mass line of the body which is responsible for a slight increase of the thickness of this region after each revolution.
Such behavior is in agreement with the numerical integration of the orbits for small values of both spin and quadrupole parameters.

Most of these features are also shared by the companion analysis performed in the case of a Schwarzschild spacetime \cite{quadrupschw}. Nevertheless, we face here with the new couplings among the spin of the background source and that of the body, thus enriching the physical content of the interaction analyzed.
Such spin-spin interactions (for each individual object itself and among them) can be fully traced at the level of the weak field limit and slow motion, as we have explicitly shown.

\appendix

\section{ZAMO relevant quantities}

We list below the non-vanishing components of the electric and magnetic parts of the Riemann tensor as well as the relevant kinematical quantities as measured by ZAMOs evaluated on the equatorial plane.

The radial components of the acceleration and expansion vectors are given by
\begin{eqnarray}
a(n)^{\hat r}&=&\frac{M}{r^2\sqrt{\Delta}}\frac{(r^2+a^2)^2-4a^2Mr}{r^3+a^2r+2a^2M}\,,\nonumber\\
\theta_\phi(n)^{\hat r}&=&-\frac{aM(3r^2+a^2)}{r^2(r^3+a^2r+2a^2M)}\,.
\end{eqnarray}
The radial components of the curvature vector are 
\begin{eqnarray}
\kappa(r,n)^{\hat r}&=&\frac{Mr-a^2}{r^2\sqrt{\Delta}}\,,\qquad
\kappa(\theta,n)^{\hat r}=-\frac{\sqrt{\Delta}}{r^2}\,,\nonumber\\
k_{\rm (lie)}&=&-\frac{(r^3-a^2M)\sqrt{\Delta}}{r^2(r^3+a^2r+2a^2M)}\,.
\end{eqnarray}
Finally, the nontrivial components of the electric and magnetic parts of the Riemann tensor with respect to ZAMOs are given by
\begin{eqnarray}
\label{E_H}
E_{\hat r \hat r}&=& -\frac{M(2 r^4+5 r^2 a^2-2 a^2 M r+3 a^4)}{r^4 (r^3+a^2 r+2 a^2 M)}\,, \nonumber\\
E_{\hat \theta \hat \theta}&=&  \frac{M (r^4+4 r^2 a^2-4 a^2 M r+3 a^4)}{r^4 (r^3+a^2 r+2 a^2 M)}
=-\frac{M}{r^3}- E_{\hat r \hat r}\,,\nonumber\\
H_{\hat r \hat \theta}&=&  -\frac{3 M a (r^2+a^2) \sqrt{\Delta}}{r^4 (r^3+a^2 r+2 a^2 M)}\,,
\end{eqnarray}
and $E_{\hat \phi \hat \phi}=-E_{\hat r \hat r}-E_{\hat \theta \hat \theta}=M/r^3$.
Note that 
\beq
\frac{E_{\hat r \hat r}}{H_{\hat r \hat \theta}}=\frac13\left(2\frac{r^2+a^2}{a\sqrt{\Delta}}+\frac{a\sqrt{\Delta}}{r^2+a^2}\right)\,.
\eeq

\section{Frame components of both spin and quadrupole force and torque}

We list below the explicit expressions of the components of both spin and quadrupole force and torque with respect to the frame adapted to $u$.

The spin force is given by Eq. (\ref{fspinframeu}) with

\begin{widetext} 

\begin{eqnarray}
\label{fspinframeu2}
F_{\rm (spin)}^1&=&s\gamma\gamma_u\left\{
\nu_u\cos2\alpha_u E_{\hat r\hat r}
+[\nu\cos(\alpha_u-\alpha)+\nu_u\cos^2\alpha_u]E_{\hat \theta\hat \theta}
+\sin\alpha_u[1+\nu\nu_u\cos(\alpha_u-\alpha)]H_{\hat r\hat \theta}
\right\}\,,\nonumber\\
F_{\rm (spin)}^2&=&-s\gamma\gamma_u^2\left\{
\nu_u[\sin2\alpha_u-\nu\nu_u\sin(\alpha_u+\alpha)]E_{\hat r\hat r}
+[\nu\sin(\alpha_u-\alpha)+\nu_u\cos\alpha_u(\sin\alpha_u-\nu\nu_u\sin\alpha)]E_{\hat \theta\hat \theta}\right.\nonumber\\
&&\left.
-[\cos\alpha_u(1+\nu\nu_u\cos(\alpha_u-\alpha))-2\nu\nu_u\cos\alpha)]H_{\hat r\hat \theta}
\right\}\,,
\end{eqnarray}
the remaining component $F_{\rm (spin)}^0$ following from the condition $F_{\rm (spin)}\cdot U=0$, i.e., 
\beq
\gamma_u(1-\nu\nu_u\cos(\alpha_u-\alpha))F_{\rm (spin)}^0=F_{\rm (spin)}^1\nu\sin(\alpha_u-\alpha)
+F_{\rm (spin)}^2\gamma_u(-\nu_u+\nu\cos(\alpha_u-\alpha))\,.
\eeq

The spin torque is instead given by Eq. (\ref{dspinframeu}) with 
\beq
\label{dspinframeu3}
{\mathcal E}(u)_{\rm (spin)}=m\gamma\left[\nu\sin(\alpha_u-\alpha)\omega^{1}
+\gamma_u(\nu\cos(\alpha_u-\alpha)-\nu_u)\omega^{2}\right]\,.
\eeq

Concerning the the quadrupole terms, the force is given by Eq. (\ref{fquadframeu}) with

\begin{eqnarray}
\label{fquadframeu3}
F_{\rm (quad)}^1&=&
\frac{2}{3}\gamma_u^2\bigg\{\sin\alpha_u\bigg[
b_1\cos2\alpha_u[(X(u)_{11}+2X(u)_{22})\nu_u^2+X(u)_{11}-X(u)_{22}]\nonumber\\
&&
-\frac{1}{\gamma_u^2}(b_2X(u)_{11}+b_3X(u)_{22})+b_2(2X(u)_{11}+X(u)_{22})\bigg]\nonumber\\
&&
+\nu_u(b_4\cos2\alpha_u+b_5)(2X(u)_{11}+X(u)_{22})
\bigg\}
\,,\nonumber\\
F_{\rm (quad)}^2&=&
\frac{2}{3}\gamma_u^3\cos\alpha_u\bigg\{
b_1\cos2\alpha_u[(X(u)_{11}+2X(u)_{22})\nu_u^2+X(u)_{11}-X(u)_{22}]\nonumber\\
&&
+\nu_u\sin\alpha_u\left[-\frac{2a_2}{\gamma_u^2}(X(u)_{11}+2X(u)_{22})+2(a_1-b_4)(2X(u)_{11}+X(u)_{22})\right]\nonumber\\
&&
-\frac{1}{\gamma_u^2}(c_1X(u)_{11}+c_2X(u)_{22})+c_3(2X(u)_{11}+X(u)_{22})
\bigg\}
\,,
\end{eqnarray}
with
\begin{eqnarray}
a_1&=&(E_{\hat \theta\hat \theta}+2E_{\hat r\hat r})\theta_{\hat\phi}(n)^{\hat r}
+H_{\hat r\hat \theta}a(n)^{\hat r}
\,,\nonumber\\
a_2&=&(2E_{\hat \theta\hat \theta}+E_{\hat r\hat r})a(n)^{\hat r}
-H_{\hat r\hat \theta}\theta_{\hat\phi}(n)^{\hat r}
\,,\nonumber\\
b_1&=&-2(E_{\hat \theta\hat \theta}+2E_{\hat r\hat r})k_{\rm (lie)}
+(2E_{\hat \theta\hat \theta}+E_{\hat r\hat r})a(n)^{\hat r}
-4H_{\hat r\hat \theta}\theta_{\hat\phi}(n)^{\hat r}
+\frac12\partial_{\hat r}E_{\hat \theta\hat \theta}
\,,\nonumber\\
b_2&=&-(E_{\hat \theta\hat \theta}+2E_{\hat r\hat r})k_{\rm (lie)}
+2H_{\hat r\hat \theta}\theta_{\hat\phi}(n)^{\hat r}
-\frac32\partial_{\hat r}E_{\hat \theta\hat \theta}
\,,\nonumber\\
b_3&=&-2(E_{\hat \theta\hat \theta}+2E_{\hat r\hat r})k_{\rm (lie)}
-2H_{\hat r\hat \theta}\theta_{\hat\phi}(n)^{\hat r}
\,,\nonumber\\
b_4&=&H_{\hat r\hat \theta}k_{\rm (lie)}
-3E_{\hat \theta\hat \theta}\theta_{\hat\phi}(n)^{\hat r}
+\partial_{\hat r}H_{\hat r\hat \theta}
\,,\nonumber\\
b_5&=&H_{\hat r\hat \theta}k_{\rm (lie)}
-(E_{\hat \theta\hat \theta}+2E_{\hat r\hat r})\theta_{\hat\phi}(n)^{\hat r}
-\partial_{\hat r}H_{\hat r\hat \theta}
\,,
\end{eqnarray}
and $c_1=c_3+2a_2$, $c_2=c_3-b_2$ and $c_3=[7b_2+4a_2+6(b_1-2b_3)]/5$, whereas $F_{\rm (quad)}^0$ can be obtained from the vanishing of the coordinate component $F_{{\rm (quad)}\,t}=0$, which implies
\begin{eqnarray}
0&=&\gamma_u(F_{\rm (quad)}^0+\nu_u F_{\rm (quad)}^2)+\frac{\sqrt{g_{\phi\phi}}N^{\phi}}{N}\left[
F_{\rm (quad)}^1\cos\alpha_u
-\gamma_u\sin\alpha_u(\nu_u F_{\rm (quad)}^0+F_{\rm (quad)}^2)
\right]\,.
\end{eqnarray}

Finally, the torque term is given by Eqs. (\ref{dquadframeu})--(\ref{dquadframeu2}) with components
\begin{eqnarray}
\label{quadtorquecompts}
{\mathcal E}(u)_{\rm (quad)}{}_1 &=&- \frac{4}{3}\gamma_u\cos\alpha_u(X(u)_{11}+2X(u)_{22})\left[
\nu_u\sin\alpha_u(2E_{\hat r\hat r}+E_{\hat \theta\hat \theta})-H_{\hat r\hat \theta}
\right]\,,\nonumber\\
{\mathcal E}(u)_{\rm (quad)}{}_2 &=&-\frac{4}{3}\gamma^2_u(2X(u)_{11}+X(u)_{22})\left[
\nu_u\cos2\alpha_uE_{\hat r\hat r}+\nu_u(1+\cos^2\alpha_u)E_{\hat \theta\hat \theta}+\sin\alpha_u(1+\nu_u^2)H_{\hat r\hat \theta}
\right]\,,\nonumber\\
{\mathcal B}(u)_{\rm (quad)}{}_3 &=& \frac{4}{3}\gamma_u\cos\alpha_u(X(u)_{11}-X(u)_{22})\left[
\sin\alpha_u(2E_{\hat r\hat r}+E_{\hat \theta\hat \theta})-\nu_uH_{\hat r\hat \theta}
\right]\,.
\end{eqnarray}

\end{widetext}

\section{Numerical integration of the full set of MPD equations}

Under the assumptions discussed in Section III on the structure of the body as well as on its motion the whole set of MPD equations (\ref{papcoreqs1})--(\ref{tulczconds}) reduces to 
\begin{eqnarray} 
\label{setfin}
\frac{\rmd m}{\rmd \tau} &=& 
F_{\rm (spin)}^0 + F_{\rm (quad)}^0\,,\nonumber\\
\frac{\rmd \alpha_u}{\rmd \tau} &=& 
-\frac{\gamma}{\nu_u}\left[\nu\cos(\alpha_u+\alpha)-\nu_u\right]\theta_{\hat\phi}(n)^{\hat r}\nonumber\\
&& 
+\frac{\gamma}{\nu_u}\left(\sin\alpha_ua(n)^{\hat r}+\nu\nu_u\sin\alpha k_{\rm (lie)}\right)\nonumber\\
&&
-\frac{1}{m\gamma_u\nu_u}(F_{\rm (spin)}^1 + F_{\rm (quad)}^1)\,,\nonumber\\
\frac{\rmd \nu_u}{\rmd \tau} &=& 
-\frac{\gamma}{\gamma_u^2}\left[\cos\alpha_ua(n)^{\hat r}+\nu\sin(\alpha_u+\alpha)\theta_{\hat\phi}(n)^{\hat r}\right]\nonumber\\
&&
+\frac{1}{m\gamma_u^2}(F_{\rm (spin)}^2 + F_{\rm (quad)}^2)\,,\nonumber\\
\frac{\rmd s}{\rmd \tau} &=&
{\mathcal B}(u)_{\rm (quad)}^3\,,\nonumber\\
\end{eqnarray}
together with the following two compatibility conditions coming from the spin evolution equations 
\begin{eqnarray} 
\label{compatib}
0&=&m({\mathcal E}(u)_{\rm (spin)}^1+{\mathcal E}(u)_{\rm (quad)}^1)+s(F_{\rm (spin)}^2 + F_{\rm (quad)}^2)\,,\nonumber\\
0&=&-m({\mathcal E}(u)_{\rm (spin)}^2+{\mathcal E}(u)_{\rm (quad)}^2)+s(F_{\rm (spin)}^1 + F_{\rm (quad)}^1)\,,\nonumber\\
\end{eqnarray}
also summarized by the vectorial relation
\beq
e_3\times({\mathcal E}(u)_{\rm (spin)}+{\mathcal E}(u)_{\rm (quad)}) +\frac{s}{m}(F_{\rm (spin)} + F_{\rm (quad)})\,.
\eeq
The evolution equation for the spin invariant implies that the quadrupolar structure of the body is responsible for the onset of spin angular momentum, if the body is initially not spinning.
Note that if $X(u)_{11}=X(u)_{22}$, as in the case of a spin-induced quadrupole, Eqs. (\ref{quadtorquecompts}) imply ${\mathcal B}(u)_{\rm (quad)}^3=0$, so that the magnitude of the spin vector remains constant.

Equations (\ref{compatib}) give two algebraic relations involving the remaining unknowns $\nu$ and $\alpha$.
After some manipulation we find
\beq
\label{nuandalpha}
\tan\alpha=\frac{A+B\gamma}{C+D\gamma}\,, \qquad
\gamma=\frac{k_1+\sqrt{k_1^2+k_2k_3}}{k_2}\,,
\eeq
where
\begin{eqnarray} 
k_1&=&AB+CD\,,\qquad
k_2=1-B^2-D^2\,,\nonumber\\
k_3&=&1+A^2+C^2\,,
\end{eqnarray}
and 

\begin{widetext} 

\begin{eqnarray} 
\label{ABC_etc}
\lambda A&=&
\left(sF_{\rm (quad)}^1-m{\mathcal E}(u)_{\rm (quad)}^2\right)
\left\{
[m^2-s^2E_{\hat r\hat r}+s^2\gamma_u^2(E_{\hat r\hat r}-E_{\hat \theta\hat \theta})]\sin\alpha_u
-s^2H_{\hat r\hat \theta}\gamma_u^2\nu_u(1+\sin^2\alpha_u)
\right\}\nonumber\\
&&
+\left(sF_{\rm (quad)}^2+m{\mathcal E}(u)_{\rm (quad)}^1\right)
[m^2-s^2E_{\hat \theta\hat \theta}-s^2H_{\hat r\hat \theta}\nu_u\sin\alpha_u]\gamma_u\cos\alpha_u
\,,\nonumber\\
B&=&\frac{(m^2-s^2E_{\hat r\hat r})\nu_u\sin\alpha_u+s^2H_{\hat r\hat \theta}}{m^2-s^2E_{\hat \theta\hat \theta}-s^2H_{\hat r\hat \theta}\nu_u\sin\alpha_u}\,,\nonumber\\
\lambda C&=&
\left(sF_{\rm (quad)}^1-m{\mathcal E}(u)_{\rm (quad)}^2\right)\cos\alpha_u
\left\{
m^2-s^2E_{\hat \theta\hat \theta}-s^2\gamma_u^2\nu_u[(E_{\hat r\hat r}+2E_{\hat \theta\hat \theta})\nu_u+H_{\hat r\hat \theta}\sin\alpha_u]
\right\}\nonumber\\
&&
-\left(sF_{\rm (quad)}^2+m{\mathcal E}(u)_{\rm (quad)}^1\right)
[m^2-s^2E_{\hat \theta\hat \theta}-s^2H_{\hat r\hat \theta}\nu_u\sin\alpha_u]\gamma_u\sin\alpha_u
\,,\nonumber\\
D&=&\frac{m^2+s^2(E_{\hat r\hat r}+E_{\hat \theta\hat \theta})}{m^2-s^2E_{\hat \theta\hat \theta}-s^2H_{\hat r\hat \theta}\nu_u\sin\alpha_u}\nu_u\cos\alpha_u\,,\nonumber\\
\lambda &=&\gamma_u[m^2-s^2E_{\hat \theta\hat \theta}-s^2H_{\hat r\hat \theta}\nu_u\sin\alpha_u]\left\{
m^2+\frac12s^2E_{\hat \theta\hat \theta}(1-3\gamma_u^2)\right.\nonumber\\
&&\left.
-s^2\gamma_u^2\nu_u\left[\frac12\nu_u\cos2\alpha_u(2E_{\hat r\hat r}+E_{\hat \theta\hat \theta})+2H_{\hat r\hat \theta}\sin\alpha_u\right]
\right\}\,.
\end{eqnarray}

\end{widetext}

Note that the following relations hold
\beq
\nu\sin\alpha=\frac{A}{\gamma}+B\,,\qquad
\nu\cos\alpha=\frac{C}{\gamma}+D\,,
\eeq
and the first equation of Eqs. (\ref{setfin}) governing the mass evolution can be written
\beq
\label{massevol}
\frac{\rmd m}{\rmd \tau} = c_1A+c_2C+ F_{\rm (quad)}^0\,,
\eeq
with
\begin{eqnarray}
c_1&=&-s^2\gamma_u\nu_u\cos\alpha_u[E_{\hat r\hat r}+2E_{\hat \theta\hat \theta}+H_{\hat r\hat \theta}\nu_u\sin\alpha_u]\,,\nonumber\\
c_2&=&-s^2\gamma_u[(E_{\hat r\hat r}-E_{\hat \theta\hat \theta})\nu_u\sin\alpha_u\nonumber\\
&&
-H_{\hat r\hat \theta}(1+\nu_u^2\sin^2\alpha_u)]\,,
\end{eqnarray}
once the dependence on $\nu$ and $\alpha$ in the component $F_{\rm (spin)}^0$ has been eliminated.
The rhs of Eq. (\ref{massevol}) vanishes for vanishing quadrupole (i.e., $F_{\rm (quad)}^a=0={\mathcal E}(u)_{\rm (quad)}^a$, so that $A=0=C$), yielding the well known constant mass result for a purely spinning particle.

Finally, in order to perform a numerical integration of Eqs. (\ref{setfin}) the evolution equations $U=\rmd x^\alpha/\rmd\tau$ must be also taken into account, i.e., 
\begin{eqnarray}
\label{Ucompts}
\frac{\rmd t}{\rmd \tau} &=&\frac{\gamma}{N}\,,\qquad
\frac{\rmd r}{\rmd \tau} =\frac{\gamma\nu\cos\alpha}{\sqrt{g_{rr}}}\,,\nonumber\\
\frac{\rmd \phi}{\rmd \tau}&=& \frac{\gamma}{\sqrt{g_{\phi\phi}}}\left(\nu\sin\alpha-\frac{\sqrt{g_{\phi\phi}}N^\phi}{N}\right)\,.
\end{eqnarray}

As a consistency check, the total energy $E$ and angular momentum $J$ given by Eq. (\ref{totalenergy}), i.e., 
\begin{eqnarray}
&&E=N\gamma_u\left\{\left[m+s(\nu_u\sin\alpha_u a(n)^{\hat r}+\theta_{\hat\phi}(n)^{\hat r})\right]\right.\nonumber\\
&&\left.
-\frac{\sqrt{g_{\phi\phi}}N^\phi}{N}\left[m\nu_u\sin\alpha_u-s(k_{\rm (lie)}+\nu_u\sin\alpha_u\theta_{\hat\phi}(n)^{\hat r})\right]
\right\}
\,,\nonumber\\
&&J=\gamma_u\sqrt{g_{\phi\phi}}\left[m\nu_u\sin\alpha_u-s(k_{\rm (lie)}+\nu_u\sin\alpha_u\theta_{\hat\phi}(n)^{\hat r})\right]
\,,\nonumber\\
\end{eqnarray}
remain constant and equal to their initial values.

\begin{widetext} 

\section{Second order solution coefficients}
\label{const}

We list below the integration constants of the second order solution (\ref{solord2}):
\begin{eqnarray} 
\label{solord2coeffs}
A_1&=&
\frac{\nu_\pm^2}{\gamma_\pm}\frac{k_{\rm (lie)}}{\Omega_{\rm(ep)}^2}(B_1\Omega_{\rm(ep)}-B_3)
+\frac{{\mathcal V}^{(r)}_{\hat s}}{M\Omega_{\rm(ep)}}\left\{
\gamma_\pm\left[\pm\gamma_\pm^3\nu_\pm^2\frac{\zeta_K}{\Omega_{\rm(ep)}}{\mathcal V}^{(r)}_{\hat s}
+(3\nu_\pm^2-2)Mk_{\rm (lie)}\pm2\nu_\pm M\zeta_K\right]\tilde{\mathcal V}^{(\phi)}_{\hat s}\right.\nonumber\\
&&\left.
+M\Omega_{\rm(ep)}\nu_\pm\left[
-2\gamma_\pm^2+\frac{1}{\Omega_{\rm(ep)}^2}\left(
(2\nu_\pm^2-3)k_{\rm (lie)}^2\mp2\nu_\pm\zeta_Kk_{\rm (lie)}+(4\nu_\pm^2+1)\gamma_\pm^2\zeta_K^2\right)\right]{\mathcal V}^{(r)}_{\hat s}\right.\nonumber\\
&&\left.
-2\gamma_\pm^3\nu_\pm^2({\mathcal V}^{(r)}_{\hat s})^2
+\frac{M^2}{\gamma_\pm}(\zeta_K^2-\Omega_{\rm(ep)}^2)
\right\}{\hat s}^2\,,\nonumber\\
A_2&=&\frac{\nu_\pm^2}{2\gamma_\pm}\frac{k_{\rm (lie)}}{\Omega_{\rm(ep)}}B_2
+({\mathcal V}^{(r)}_{\hat s})^2\left\{
\frac{\gamma_\pm^3\nu_\pm^2}{2M\Omega_{\rm(ep)}}\left({\mathcal V}^{(r)}_{\hat s}
\mp2\gamma_\pm\frac{\zeta_K}{\Omega_{\rm(ep)}}\tilde{\mathcal V}^{(\phi)}_{\hat s}\right)
+(3\nu_\pm^2-1)\frac{\gamma_\pm^2}{4\nu_\pm}
\right.\nonumber\\
&&\left.
-\frac{1}{4\Omega_{\rm(ep)}^2}\left[
(2\nu_\pm^2-3)k_{\rm (lie)}^2\mp2\nu_\pm\zeta_Kk_{\rm (lie)}+(4\nu_\pm^2+1)\gamma_\pm^2\zeta_K^2\right]{\mathcal V}^{(r)}_{\hat s}
\right\}{\hat s}^2
\equiv{\hat s}^2A_{2{\hat s}}\,,\nonumber\\
A_3&=&
-\frac{\nu_\pm^2}{\gamma_\pm}\frac{k_{\rm (lie)}}{\Omega_{\rm(ep)}}B_3
\equiv{\hat s}^2A_{3{\hat s}}
\,,\nonumber\\
B_1&=&
-2B_2-\frac{B_3}{\Omega_{\rm(ep)}}-\frac{1}{\nu_\pm\Omega_{\rm(ep)}}\left[
\pm2\gamma_\pm\zeta_KA_4+\frac{1}{m\gamma_\pm}\left(F^1_{\rm (quad)}\pm\zeta_K{\mathcal E}(u)_{\rm (quad)}^2\right)\right]\nonumber\\
&&
+\left\{
\frac{\gamma_\pm}{\nu_\pm M\Omega_{\rm(ep)}}\left[
M(k_{\rm (lie)}\mp4\gamma_\pm^2\nu_\pm\zeta_K)\tilde{\mathcal V}^{(\phi)}_{\hat s}
-M^2(7\zeta_K^2-2\Omega_{\rm(ep)}^2)
\right]\tilde{\mathcal V}^{(\phi)}_{\hat s}\right.\nonumber\\
&&\left.
+\frac{1}{\nu_\pm}\left[
\frac{\gamma_\pm^2}{\nu_\pm}\pm M\zeta_K+\frac{4\gamma_\pm^2}{3M\Omega_{\rm(ep)}^2}(\nu_\pm k_{\rm (lie)}\mp3\zeta_K)({\mathcal V}^{(r)}_{\hat s})^2
\right]{\mathcal V}^{(r)}_{\hat s}
\right\}{\hat s}^2
\,,\nonumber\\
B_2&=&
({\mathcal V}^{(r)}_{\hat s})^2\left\{
\frac{\gamma_\pm^2}{3\nu_\pm M\Omega_{\rm(ep)}^2}(\nu_\pm k_{\rm (lie)}\mp3\zeta_K)\left({\mathcal V}^{(r)}_{\hat s}
\pm2\gamma_\pm\frac{\zeta_K}{\Omega_{\rm(ep)}}\tilde{\mathcal V}^{(\phi)}_{\hat s}\right)
-\frac{\gamma_\pm}{2\nu_\pm}\frac{k_{\rm (lie)}}{\Omega_{\rm(ep)}}
\right.\nonumber\\
&&\left.
+\frac{2\zeta_K^2}{\gamma_\pm\nu_\pm\Omega_{\rm(ep)}^3}(k_{\rm (lie)}\mp2\gamma_\pm^2\nu_\pm\zeta_K)
-\frac{\gamma_\pm^3}{6\nu_\pm\Omega_{\rm(ep)}^3}\left[\nu_\pm^2\partial_{\hat r}E_{\hat \theta\hat \theta}+2\nu_\pm\partial_{\hat r}H_{\hat r\hat \theta}-\partial_{\hat r}E_{\hat r\hat r}\right]
\right\}{\hat s}^2
\equiv{\hat s}^2B_{2{\hat s}}
\,,\nonumber\\
B_3&=&-3\Omega_{\rm(ep)}B_2
+\frac{{\mathcal V}^{(r)}_{\hat s}}{\nu_\pm}\left\{
\frac{2\gamma_\pm^2}{M\Omega_{\rm(ep)}}(\nu_\pm k_{\rm (lie)}\mp3\zeta_K)({\mathcal V}^{(r)}_{\hat s})^2
+\frac{\gamma_\pm^3}{M}\left({\mathcal V}^{(r)}_{\hat s}
\mp2\gamma_\pm\frac{\zeta_K}{\Omega_{\rm(ep)}}\tilde{\mathcal V}^{(\phi)}_{\hat s}\right){\mathcal V}^{(r)}_{\hat s}\tilde{\mathcal V}^{(\phi)}_{\hat s}\right.\nonumber\\
&&\left.
-\frac32\gamma_\pm k_{\rm (lie)}{\mathcal V}^{(r)}_{\hat s}
\pm\frac{3\zeta_K}{\Omega_{\rm(ep)}}(k_{\rm (lie)}\mp2\gamma_\pm^2\nu_\pm\zeta_K)\tilde{\mathcal V}^{(\phi)}_{\hat s}
+\frac{M\zeta_K^2}{\Omega_{\rm(ep)}}(5\nu_\pm k_{\rm (lie)}\mp2\zeta_K)
-M\Omega_{\rm(ep)}(\nu_\pm k_{\rm (lie)}\mp\zeta_K)\right.\nonumber\\
&&\left.
+\frac{M\gamma_\pm^2}{2\Omega_{\rm(ep)}}\left[(\nu_\pm^2+1)\partial_{\hat r}H_{\hat r\hat \theta}-\nu_\pm\partial_{\hat r}(E_{\hat r\hat r}-E_{\hat \theta\hat \theta})\right]
\right\}{\hat s}^2
\equiv{\hat s}^2B_{3{\hat s}}
\,,\nonumber\\
C_1&=&
\gamma_\pm\frac{\sqrt{\Delta}}{r_0\Omega_{\rm(ep)}^2}\left\{
\nu_\pm(B_1\Omega_{\rm(ep)}-B_3)
+\gamma_\pm{\mathcal V}^{(r)}_{\hat s}\left[
\left(\kappa(r,n)^{\hat r}+k_{\rm (lie)}\right){\mathcal V}^{(r)}_{\hat s}
-\frac{\gamma_\pm}{\nu_\pm}\Omega_{\rm(ep)}\tilde{\mathcal V}^{(\phi)}_{\hat s}
\right]{\hat s}^2\right\}\,,\nonumber\\
C_2&=&
\gamma_\pm\frac{\sqrt{\Delta}}{2r_0\Omega_{\rm(ep)}^2}\left\{
\nu_\pm B_2
-\frac{\gamma_\pm}{2\Omega_{\rm(ep)}}\left(\kappa(r,n)^{\hat r}+k_{\rm (lie)}\right)({\mathcal V}^{(r)}_{\hat s})^2{\hat s}^2\right\}
\equiv{\hat s}^2C_{2{\hat s}}
\,,\nonumber\\
C_3&=&
-\gamma_\pm\nu_\pm\frac{\sqrt{\Delta}}{r_0\Omega_{\rm(ep)}^2}B_3
\equiv{\hat s}^2C_{3{\hat s}}
\,,\nonumber\\
D_1&=&
\mp2\gamma_\pm^2\nu_\pm^2\frac{\zeta_K}{N\Omega_{\rm(ep)}^3}(B_1\Omega_{\rm(ep)}-2B_3)
+\frac{{\mathcal V}^{(r)}_{\hat s}}{MN\Omega_{\rm(ep)}^2}\left\{
-2\gamma_\pm^2\left[\pm2\gamma_\pm^3\nu_\pm\frac{\zeta_K}{\Omega_{\rm(ep)}}{\mathcal V}^{(r)}_{\hat s}
+\nu_\pm Mk_{\rm (lie)}\mp\gamma_\pm^2(1+2\nu_\pm^2)M\zeta_K\right]\tilde{\mathcal V}^{(\phi)}_{\hat s}\right.\nonumber\\
&&\left.
+M\Omega_{\rm(ep)}\gamma_\pm^3\left[
1\mp\frac{2\nu_\pm\zeta_K}{\Omega_{\rm(ep)}^2}\left(
(3\nu_\pm^2-4)k_{\rm (lie)}\pm6\nu_\pm\zeta_K\right)\right]{\mathcal V}^{(r)}_{\hat s}
+2\gamma_\pm^4\nu_\pm({\mathcal V}^{(r)}_{\hat s})^2
+\gamma_\pm^2\nu_\pm M^2(\zeta_K^2-\Omega_{\rm(ep)}^2)
\right\}{\hat s}^2\,,\nonumber\\
D_2&=&
\mp\gamma_\pm^2\nu_\pm^2\frac{\zeta_K}{2N\Omega_{\rm(ep)}^2}B_2\nonumber\\
&&
-\frac{\gamma_\pm^3}{4N\Omega_{\rm(ep)}^2}({\mathcal V}^{(r)}_{\hat s})^2\left\{
\frac{\gamma_\pm\nu_\pm^2}{4M\Omega_{\rm(ep)}}\left({\mathcal V}^{(r)}_{\hat s}
\mp2\gamma_\pm\frac{\zeta_K}{\Omega_{\rm(ep)}}\tilde{\mathcal V}^{(\phi)}_{\hat s}\right)
+1\mp\frac{\nu_\pm\zeta_K}{\Omega_{\rm(ep)}^2}\left[
(3\nu_\pm^2-4)k_{\rm (lie)}\pm6\nu_\pm\zeta_K\right]
\right\}{\hat s}^2
\equiv{\hat s}^2D_{2{\hat s}}
\,,\nonumber\\
D_3&=&
\mp2\gamma_\pm^2\nu_\pm^2\frac{\zeta_K}{N\Omega_{\rm(ep)}^2}B_3
\equiv{\hat s}^2D_{3{\hat s}}
\,,\nonumber\\
D_4&=&
\frac{\gamma_\pm^3}{2N}\left[2\nu_\pm A_4+\gamma_\pm^2(1+2\nu_\pm^2)(\tilde{\mathcal V}^{(\phi)}_{\hat s})^2{\hat s}^2\right]
-\Omega_{\rm(ep)}(D_1+2D_2)-D_3
\,,\nonumber\\
E_1&=&
\bar\zeta_\pm D_1
-\frac{\gamma_\pm}{\nu_\pm\sqrt{g_{\phi\phi}}\Omega_{\rm(ep)}}({\mathcal V}^{(r)}_{\hat s})^2{\hat s}^2\,,\nonumber\\
E_2&=&
\bar\zeta_\pm D_2
+\frac{\gamma_\pm}{4\nu_\pm\sqrt{g_{\phi\phi}}\Omega_{\rm(ep)}}({\mathcal V}^{(r)}_{\hat s})^2{\hat s}^2
\equiv{\hat s}^2E_{2{\hat s}}
\,,\nonumber\\
E_3&=&
\bar\zeta_\pm D_3
\equiv{\hat s}^2E_{3{\hat s}}
\,,\nonumber\\
E_4&=&
\bar\zeta_\pm D_4
+\frac{\gamma_\pm}{2\nu_\pm\sqrt{g_{\phi\phi}}}\left[({\mathcal V}^{(r)}_{\hat s})^2-\gamma_\pm^2(\tilde{\mathcal V}^{(\phi)}_{\hat s})^2\right]{\hat s}^2\,,
\end{eqnarray}
with 
\begin{eqnarray}
F^1_{\rm (quad)}&=&-\frac23\gamma_\pm^2\left[
(1+\nu_\pm^2)\partial_{\hat r}E_{\hat \theta\hat \theta}+2\nu_\pm\partial_{\hat r}H_{\hat r\hat \theta}
\right](2{\hat X}_{11}+{\hat X}_{22})
-\frac{4\gamma_\pm\Omega_{\rm(ep)}\tilde{\mathcal V}^{(r)}_{\hat s}}{M(1+\nu_\pm^2)}(\nu_\pm k_{\rm (lie)}\pm\zeta_K)[({\hat X}_{11}+{\hat X}_{22})\nu_\pm^2+{\hat X}_{11}]\nonumber\\
&&
+\frac23\left[2E_{\hat \theta\hat \theta}+E_{\hat r\hat r}+\frac{E_{\hat r\hat r}-E_{\hat \theta\hat \theta}}{1+\nu_\pm^2}\right](k_{\rm (lie)}\mp2\gamma_\pm^2\nu_\pm \zeta_K)
[({\hat X}_{11}+2{\hat X}_{22})\nu_\pm^2+{\hat X}_{11}-{\hat X}_{22}]
\,,\nonumber\\
{\mathcal E}(u)_{\rm (quad)}^2&=&-\frac{4}{3}mM\Omega_{\rm(ep)}\gamma_\pm\tilde{\mathcal V}^{(r)}_{\hat s}(2{\hat X}_{11}+{\hat X}_{22})\,.
\end{eqnarray}
Note that only the coefficients $A_1,B_1,C_1,C_4,D_1,D_4,E_1,E_4$ explicitly depend on the quadrupole parameters.

\section*{Acknowledgements}
We are indebted to Profs. T. Damour and G. Faye for useful discussions at the beginning of the present project.

\end{widetext}

\end{document}